\documentclass[amsmath,amssymb,aps,prd,11pt,tightenlines,superscriptaddress,
nofootinbib,preprintnumbers,notitlepage]{revtex4-1}

\usepackage{multirow}
\usepackage{graphicx}
\usepackage{tabularx}
\usepackage{mathtools}
\usepackage{slashed}
\usepackage{url}
\usepackage{hyperref}
\usepackage[maxfloats=100]{morefloats}
\usepackage{color,xcolor}
\usepackage[normalem]{ulem}
\definecolor{dark-red}{rgb}{0.8,0.0,0.0}
\definecolor{dark-blue}{rgb}{0.0,0.0,0.8}
\definecolor{dark-green}{rgb}{0.,0.6,0.}

\makeatletter
\def\l@subsubsection#1#2{}
\makeatother

\newcommand{\toolfont}[1]{\texttt{#1}}

\newcommand{\qqquad}{\qquad \qquad}

\newcommand{\lgr}[1]{\mathcal{L}_\text{#1}}
\newcommand{\ord}[1]{\mathcal{O}\!\left(#1\right)}
\newcommand{\ope}[1]{\mathcal{O}_{#1}}
\newcommand{\ie}{i.\,e.\ }
\newcommand{\eg}{e.\,g.\ }

\newcommand{\ifb}{\ensuremath \mathrm{fb}^{-1}}
\newcommand{\gev}{{\ensuremath \mathrm{GeV}}}
\newcommand{\tev}{{\ensuremath \mathrm{TeV}}}

\newcommand{\phisq}{\phi^\dagger \phi}

\newcommand{\pder}[2]{\frac {\partial #1} {\partial #2}}

\newcommand{\fourvec}[4]{\begin{pmatrix*}[r] #1 \\ #2 \\ #3 \\ #4 \end{pmatrix*}}
\newcommand{\fourvecc}[4]{\begin{pmatrix*}[c] #1 \\ #2 \\ #3 \\ #4 \end{pmatrix*}}
\newcommand{\fivevec}[5]{\begin{pmatrix*}[r] #1 \\ #2 \\ #3 \\ #4 \\ #5 \end{pmatrix*}}
\newcommand{\fivevecc}[5]{\begin{pmatrix*}[c] #1 \\ #2 \\ #3 \\ #4 \\ #5 \end{pmatrix*}}

\newcommand{\twomatc}[4]{\begin{pmatrix*}[c] #1 & #2\\ #3 & #4 \end{pmatrix*}}

\newcommand{\overbar}[1]{\mkern 1.5mu\overline{\mkern-1.5mu#1\mkern-1.5mu}\mkern 1.5mu}

\DeclareMathOperator{\Pois}{Pois}

\newcommand{\boldx}{\ensuremath \mathbf{x}}
\newcommand{\boldv}{\ensuremath \mathbf{v}}
\newcommand{\boldg}{\ensuremath \mathbf{g}}
\newcommand{\boldzero}{\ensuremath \mathbf{0}}

\newcommand{\arxiv}[1]{\href{http://arxiv.org/abs/#1}{arXiv:#1}}

\usepackage{enumitem}
\setlist[itemize]{itemsep=1pt,parsep=1pt, topsep=1pt}

\begin{document}


\title{Better Higgs Measurements Through Information Geometry} 

\author{Johann Brehmer}
\affiliation{Institut f\"ur Theoretische Physik, Universit\"at Heidelberg, Germany}

\author{Kyle Cranmer}
\affiliation{Center for Cosmology \& Particle Physics, New York University, USA}

\author{Felix Kling}
\affiliation{University of California, Irvine, USA}
 
\author{Tilman Plehn}
\affiliation{Institut f\"ur Theoretische Physik, Universit\"at Heidelberg, Germany}

\date{\today}

\preprint{UCI-HEP-TR-2016-24}

\begin{abstract}
  Information geometry can be used to understand and optimize Higgs
  measurements at the LHC. The Fisher information encodes the maximum
  sensitivity of observables to model parameters for a given
  experiment. Applied to higher-dimensional operators, it defines the
  new physics reach of any LHC signature.  We calculate the Fisher
  information for Higgs production in weak boson fusion with decays
  into tau pairs and four leptons, and for Higgs production in
  association with a single top quark.  In a next step we analyze how
  the differential information is distributed over phase space, which
  defines optimal event selections. Conversely, we consider the
  information in the distribution of a subset of the kinematic
  variables, showing which production and decay observables are the
  most powerful and how much information is lost in traditional
  histogram-based analysis methods compared to fully multivariate
  ones.
\end{abstract}

\maketitle
\tableofcontents

\section{Introduction}
\label{sec:intro}

After its experimental discovery~\cite{higgs,discovery} the Higgs
boson and its properties have immediately become one of the most
important and active fields of searches for physics beyond the
Standard Model at the LHC.  In the Lagrangian language of fundamental
physics, the Higgs properties can be described by a continuous and
high-dimensional parameter space, for instance in terms of Wilson
coefficients in an effective field theory
(EFT)~\cite{eftfoundations,eftorig,eftreviews}. One of the main
features of moving from simple coupling modifications to
higher-dimensional operators is that we can now include kinematic
distributions in these searches~\cite{higgs_fit,yr4}. A common
challenge of all Higgs analyses is how to navigate the vast family of
phase-space distributions.

Responding to the overwhelming amount of search strategies, we expect
the LHC collaborations to focus more and more on high-level
statistical tools, including hypothesis tests based on multivariate
analysis with machine learning or the matrix element
method~\cite{statistics,kyle_review}. Historically, these tools
compare two discrete hypotheses, and applying them to continuous,
high-dimensional parameter spaces is computationally expensive.  Only
recently, machine learning techniques have been extended to include
inference on such continuous high-dimensional parameter
spaces~\cite{machine_learning}.  With these capabilities, it becomes
increasingly important to be able to effectively characterize the
information contained in these distributions.  We present an approach
based on information geometry~\cite{information-geometry},
intrinsically designed to study continuous parameter spaces of
arbitrary dimensionality without the need for any discretization of
the hypothesis. We use this to compare and improve Higgs measurement
strategies.

Our central object is the Fisher information matrix. Through the
Cram\'er-Rao bound it determines the maximum knowledge on model
parameters that we can derive from an
observation~\cite{cramer-rao,information-applications}. In that sense the
Cram\'er-Rao bound for the Fisher information plays a similar role as
the Neyman-Pearson lemma~\cite{neyman-pearson} plays for a discrete
hypothesis test and the log-likelihood ratio: it allows us to define
and to compute the best possible outcome of any multivariate black-box
analysis~\cite{kyle_review,madmax1}. In addition, the Fisher
information matrix defines a metric in the space of model parameters,
which not only provides an intuitive geometric picture, but also gives
us a handle on the linearization of the observable in terms of new
physics effects.

When we apply our information geometry framework to Higgs physics, in
particular to analyses of the dimension-6 Higgs Lagrangian, we can
tackle questions of the kind:
\begin{itemize}[label=\raisebox{0.1ex}{\scriptsize$\bullet$}]
\item What is the maximum precision with which we can measure
  continuous model parameters?
\item How is the information distributed over phase space?
\item How much of the full information is contained in a given set of
  distributions?
\item Which role do higher-dimensional corrections in the EFT
  expansion play?
\end{itemize}
\bigskip

We demonstrate our approach using three examples: Higgs production in
weak boson fusion (WBF)~\cite{dave_thesis} with its tagging-jet
kinematics~\cite{tagging} is well known to probe many aspects of the
Higgs-gauge coupling structure~\cite{phi_jj}. Focusing on the WBF
production kinematics we first analyze its combination with a Higgs
decay to tau leptons~\cite{wbf_tau}. This will for example allow us to
estimate how much of the entire information on higher-dimensional
operators is typically included in the leading tagging jet
distributions. Combining WBF production with a Higgs decay to $ZZ^*$
pairs we can test how much additional information is included in the
decay distributions. Conceptually, this contrasts two ways to
constrain the same effective Lagrangian via large momentum flow
through the relevant vertices or via precision observables~\cite{higgs_fit}.
Finally, we will test how useful Higgs production in association with
a single top~\cite{top_higgs} is for a dimension-6 operator analysis.

In a set of appendices we give a worked-out simple example for our
approach, explain how we compute the Fisher information, show more
information on our example processes, indicate how systematic or
theory uncertainties can be included, and discuss the relation of our
approach to standard log-likelihood ratios.

\section{Information geometry and Cram\'er-Rao bound}
\label{sec:formalism}

At the LHC, we typically use a set of possibly correlated event rates
$\boldx$ to measure a set of model parameters. Those can, for example,
be a vector of Higgs couplings with the unknown true value
$\boldg$. These Higgs couplings define a continuous, high-dimensional
model space.  The outcome of the measurement is an estimator for the
couplings $\hat{\boldg}$ that follows a probability distribution
$f( \hat{\boldg} | \boldg)$. For an unbiased estimator its expectation
value is equal to its true value,
\begin{align}
  \bar{g}_i 
  \equiv E \left[ \hat{g}_i \middle | \boldg \right] = g_i \,.
\end{align}
Our argument can be trivially extended to biased estimators.  The variance, or for more
than one model parameter the covariance matrix
\begin{align}
  C_{ij}(\boldg)
  \equiv E \left[ (\hat{g}_i-\bar{g}_i) (\hat{g}_j-\bar{g}_j)  \middle | \boldg \right] \,,
  \label{eq:def_cov}
\end{align}
provides a measure of the precision of the measurement.  For a set of
uncorrelated measurements the covariance matrix is a diagonal matrix
made of the individual variances.

The relation $f(\boldx|\boldg)$ between the measurement $\boldx$ and
assumed true parameters $\boldg$ can be extracted from Monte-Carlo and
detector simulations. If we know it, we can describe the reach of a
measurement using the Fisher information matrix
\begin{align}
  I_{ij}(\boldg)
     \equiv 
      - E \left[
      \frac {\partial^2 \log f(\boldx |\boldg) } {\partial g_i \, \partial g_j}  \middle | \boldg   \right] \,.
  \label{eq:fisher_information}
\end{align}
The Cram\'er-Rao bound~\cite{cramer-rao}
states that the covariance matrix in Eq.\;\eqref{eq:def_cov} is bounded
from below by the inverse Fisher information: the smallest achievable
uncertainty is then given by
\begin{align}
  C_{ij} \geq (I^{-1})_{ij} \,.
\end{align}
Large entries in the Fisher information indicate directions in model
space which can be measured well. Eigenvectors with eigenvalue zero
are blind directions.  Fortunately, the Fisher information is invariant
under a reparametrization of the observables $\boldx$, and transforms
covariantly under a reparametrization of the model parameters
$\boldg$.

After removing blind directions, the Fisher information is a symmetric
and positive definite rank-two tensor and defines a Riemannian metric
on the model space~\cite{information-geometry}. This allows us define
a local as well a as global distance measure in model space,
\begin{align}
d_\text{local}( \boldg_b; \boldg_a ) 
&= \sqrt{(\boldg_a - \boldg_b)_i \, I_{ij}(\boldg_a) \, (\boldg_a - \boldg_b)_j}
\notag \\
d(\boldg_b, \boldg_a)
&= \min_{\boldg(s)} \;
   \int_{s_a}^{s_b} d s \; \sqrt{\frac{dg_i (s)}{ds} \, I_{ij}(\boldg(s)) \, \frac{dg_j (s)}{ds}} \,,
\label{eq:distances}
\end{align}
where the global distance is the length of a geodesic (the curve that
minimizes the distance). Contours of constant distances define optimal
error ellipsoids. The distance tracks how (un-)likely it is to measure
$\hat{\boldg} = \boldg_b$ given the true value $\boldg = \boldg_a$. If
the estimator is distributed according to a multivariate Gaussian
around the true value, the local distance values directly correspond
to the difference in $\hat{\boldg}$ and $\boldg_a$ measured in
standard deviations.\bigskip

A typical LHC measurement includes an observed total number of events
$n$, distributed over possible phase space positions $\boldx$. The
probability distribution in Eq.\;\eqref{eq:fisher_information}
factorizes~\cite{kyle_review,madmax1}
\begin{align}
  f(\boldx_1, \dots, \boldx_n | \boldg) = \Pois ( n | L \sigma(\boldg) ) \; \prod_{i=1}^n f^{(1)} (\boldx_i|\boldg) \, ,
\end{align}
where $f^{(1)} (\boldx|\boldg)$ is the normalized probability
distribution for a single event populating the phase space position
$\boldx$. This can be calculated for example with Monte-Carlo
simulations. The total cross section is $\sigma(\boldg)$, to be
multiplied with the integrated luminosity $L$.  The corresponding
Fisher information is
\begin{align}
  I_{ij} 
  &= \frac{L}{\sigma} \; \pder {\sigma}{g_i}  \, \pder {\sigma}{g_j}
  - L \, \sigma \; 
    E \left[ \frac{\partial^2 \log f^{(1)}(\boldx|\boldg)}{\partial g_i \, \partial g_j} \right] \,.
\label{eq:fisher_rates}
\end{align}
The Fisher information is additive when we combine phase-space
regions. After integrating over the entire phase space, this full
Fisher information defines the minimum covariance matrix
possible.\bigskip

Instead of integrating over the entire phase space, it is enlightening
to study how the information is distributed in phase space. Consider
the differential quantity $d I_{ij} /d \boldv$, where $\boldv$ is a
kinematic variable like an invariant mass or angle calculated from
$\boldx$. Here $I_{ij}$ uses all the information in $\boldx$, but we
are able to study the distribution of the information with respect to
$\boldv$.\footnote{This is similar to how the log-likelihood ratio was
  studied differentially with
  \toolfont{MadMax}~\cite{madmax1,madmax2}.} Such a distribution
defines the important phase-space region for a measurement and should
drive the design of event selections: it allows us to calculate the
information loss from kinematic cuts, and to quantify the trade-off
between signal purity and maximal information. Integrating over this
differential information will reproduce the total Fisher information
$I_{ij}$.

Conversely, if in Eq.\;\eqref{eq:fisher_rates} we replace the full
phase space point $\boldx$ with a lower-dimensional set of kinematic
variables $\boldv$, we will arrive at the information in this reduced
set of kinematic variables.  We refer to this as the information in
distributions, and we will use this definition to identify how
efficient analyses in terms of a small number of available kinematic
distributions can be.\bigskip

For our analysis we parametrize the Higgs properties in terms of
dimension-6 operators~\cite{eftfoundations,eftorig,eftreviews},
\begin{align}
  \lgr{} = \lgr{SM} + \frac {f_i} {\Lambda^2} \ope{i} 
\quad \text{implying} \quad 
   g_i = \frac {f_i \, v^2} {\Lambda^2}  \,,
\label{eq:def_wilson}
\end{align}
where the additional factor $v^2$ ensures that our parameters $\boldg$
are dimensionless, and the Standard Model corresponds to
$\boldg = \boldzero$. In Eq.\;\eqref{eq:information_from_events}
we see that the Fisher information
around this point, $I_{ij} (\boldzero)$, only measures the linear
terms in $\boldg \propto 1/\Lambda^2$ and is not sensitive to higher
corrections.  Dimension-6 squared contributions appear away from the
Standard Model point and in the corresponding global distances.  The
difference between local and global distances thus provides a measure
of the impact of $1/\Lambda^4$ contributions~\cite{eft-edge}.

\section{Weak-boson-fusion Higgs to taus}
\label{sec:wbf_taus}

\begin{figure}
  \includegraphics[width=0.36 \textwidth]{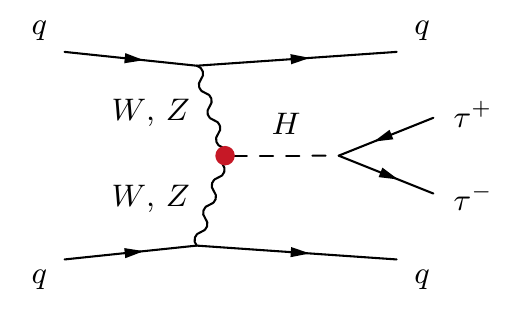}%
  \caption{Example Feynman diagram for weak-boson-fusion Higgs
    production with $H \to \tau \tau$. The red dot shows the
    Higgs-gauge interactions affected by the dimension-6 operators of
    our analysis.}
\label{fig:wbf_tautau_diag}
\end{figure}

The first question we tackle with our information geometry approach is
what we can learn about higher-dimensional operators from the
non-trivial kinematics of weak-boson-fusion production. As a decay we
include a simple fermionic two-body decay
$H \to \tau \tau$~\cite{wbf_tau}, see Fig.~\ref{fig:wbf_tautau_diag}.
For our proof of concept we stick to a parton-level analysis at
leading order.  The dominant irreducible backgrounds are QCD $Zjj$
production and electroweak $Zjj$ production, both with the decay
$Z \to \tau \tau$, and Higgs production in gluon fusion with
$H \to \tau \tau$. 

We do not simulate tau decays, but multiply the rates with the
branching ratios for the semi-leptonic di-tau mode and assume the
di-tau system can be reconstructed with the collinear approximation
with a realistic resolution for $m_{\tau\tau}$. Following the
procedure outlined in Refs.~\cite{madmax1,madmax2}, we smear the
$m_{\tau \tau}$ distributions using a Gaussian (with width 17 GeV) for
Higgs production and a double Gaussian (where the dominant component
has a width of 13~GeV) for $Z$ production, estimated from Fig.~1a of
Ref.~\cite{Aad:2015vsa}. The double Gaussian ensures an accurate
description of the high-mass tail of the $Z$ peak around
$m_{\tau\tau} = m_H$.  Otherwise, no detector effects are included.
We require loose cuts
\begin{alignat}{3}
  p_{T,j} &> 20 \ \gev  \qquad & \qquad |\eta_{j}| &< 5.0  \qquad & \qquad 
  \Delta \eta_{jj} &> 2.0  \notag \\ 
  p_{T,\tau} &> 10 \ \gev  \qquad & \qquad |\eta_{\tau}| &< 2.5 \, ,
  \label{eq:wbf_tautau_acceptance_cuts}
\end{alignat}
to include as much of phase space as possible.

The different QCD
radiation patterns of electroweak and QCD signal and background
processes are a key feature to separate the signal from the
background~\cite{tagging}. We take it into account through approximate
individual jet veto survival probabilities~\cite{wbf_tau},
\begin{align}
  \varepsilon^\text{CJV}_{\text{WBF $H$}} = 0.71 \qqquad
  \varepsilon^\text{CJV}_{\text{EW $Z$}} = 0.48 \qqquad
  \varepsilon^\text{CJV}_{\text{QCD $Z$}} = 0.14 \qqquad
  \varepsilon^\text{CJV}_{\text{GF $H$}} = 0.14 \,.
\end{align}
Since our phase space $\boldx$ does not include any jets other than
the two tagging jets, we are not sensitive to details of the central
jet veto other than the relative survival probabilities and possible
second-order effects that would correlate the veto with
$\boldx$. After the event selection of
Eq.\;\eqref{eq:wbf_tautau_acceptance_cuts} and applying the CJV
efficiencies, the WBF Higgs signal of 53~fb in the SM faces a
dominant QCD $Z$ background of 2.7~pb.

We consider five $CP$-even dimension-6 operators in the HISZ
basis~\cite{hisz,higgs_fit},
\begin{alignat}{2}
  \ope{B}  &= i \frac{g}{2} \, (D^\mu\phi^\dagger) (D^\nu\phi) \, B_{\mu\nu} \quad & \quad
  \ope{W}  &= i \frac{g}{2} \, (D^\mu\phi)^\dagger \sigma^k ( D^\nu\phi) \, W_{\mu\nu}^k \notag \\
  \ope{BB}  &= -\frac{g'^2}{4}  \,  (\phisq) \, B_{\mu\nu} \, B^{\mu\nu} \quad & \quad
  \ope{WW}  &= -\frac{g^2}{4} \, (\phisq) \, W^k_{\mu\nu} \, W^{\mu\nu\, k} \notag \\
  \ope{\phi,2}  &= \frac{1}{2} \, \partial^\mu(\phi^\dagger\phi) \, \partial_\mu(\phi^\dagger\phi) \,.
\label{eq:wbf_ope}
\end{alignat}
The first four operators introduce new Lorentz structures into
Higgs-gauge interactions, which translate into changed kinematic
shapes.  The pure Higgs operator $\ope{\phi,2}$ leads to a universal
rescaling of all single-Higgs couplings and otherwise only affects the
Higgs self-coupling. Other operators that contribute to WBF Higgs
production are tightly constrained by electroweak precision data or
can be removed from the basis using field
redefinitions~\cite{power-to-the-data}.  The effect of $\ope{\phi,2}$
on gluon-fusion Higgs production is taken into account in our
analysis, while the effects from $\ope{W}$ and $\ope{B}$ on the
subleading electroweak $Zjj$ background are neglected.

\subsection{Maximum precision on Wilson coefficients}

\begin{figure}[b]
  \includegraphics[height=0.34 \textwidth,clip,trim=0.3cm 0 0.05cm 0]{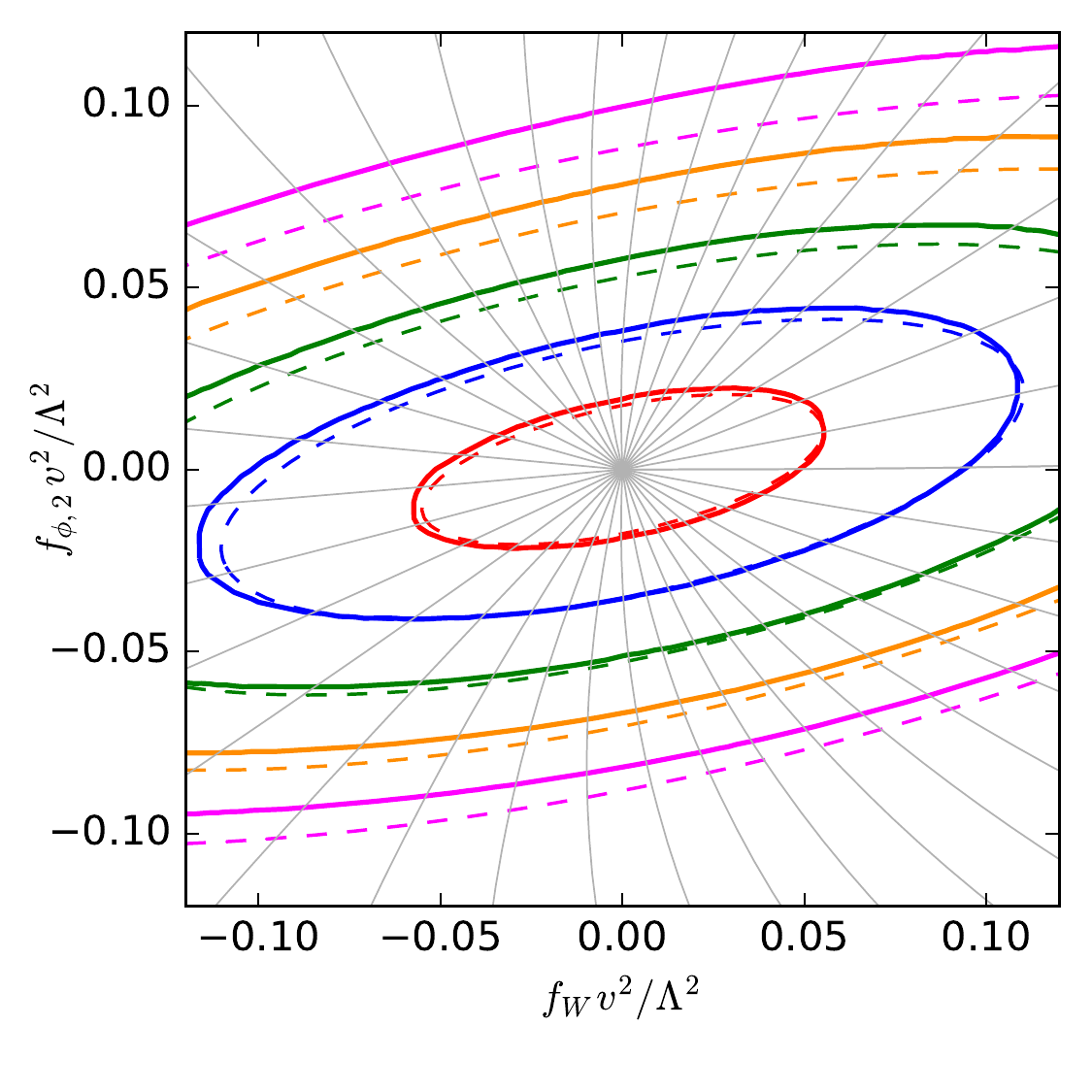}%
  \includegraphics[height=0.34 \textwidth,clip,trim=0.3cm 0 0.05cm 0]{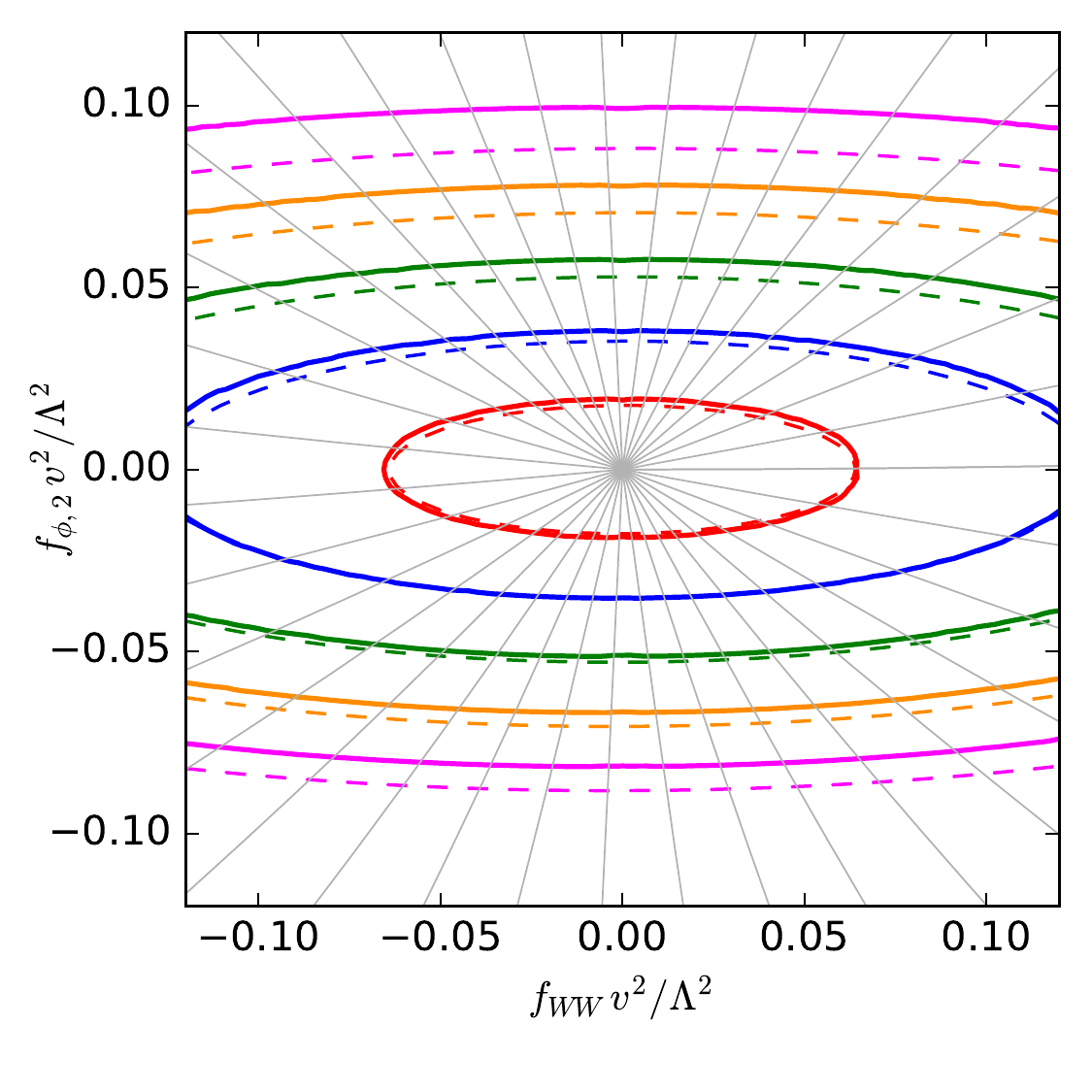}%
  \includegraphics[height=0.34 \textwidth,clip,trim=0.3cm 0 0.05cm 0]{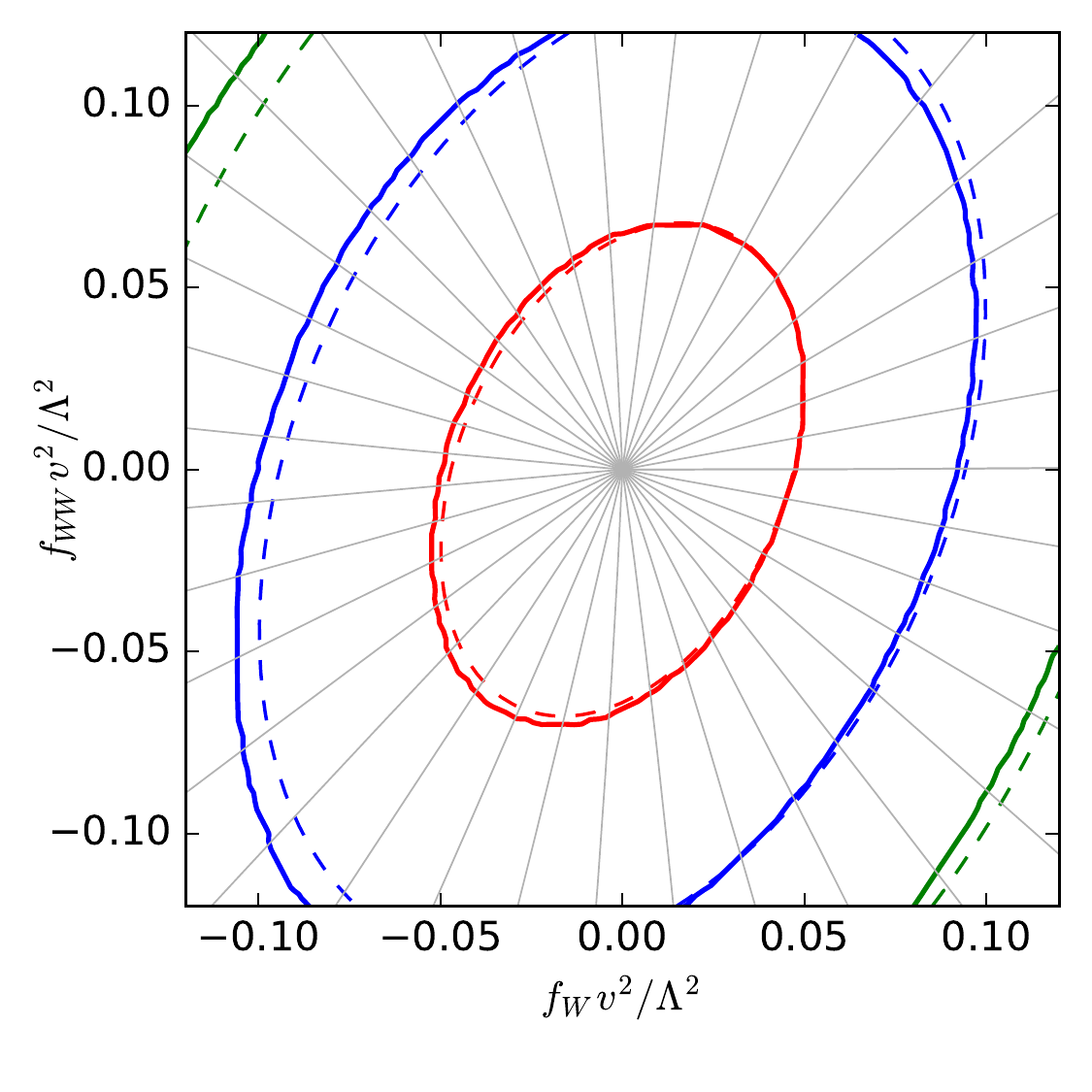}%
  \caption{Error ellipses defined by the Fisher information in the WBF
    $H \to \tau \tau$ channel. We show contours of local distance
    $d_\text{local}(\boldg ; \boldzero)$ (dashed) and global distance
    $d(\boldg,\boldzero)$ (solid).  The colored contours indicate
    distances of $d = 1~...~5$. In grey we show example geodesics. The
    $g_i$ not shown are set to zero. }
\label{fig:wbf_tautau_geometry}
\end{figure}

Following Eq.\;\eqref{eq:def_wilson}, our model space is spanned by five
dimensionless parameters
\begin{align}
  \boldg = \frac {v^2} {\Lambda^2}  \fivevecc {f_{\phi,2}} {f_W} {f_{WW}} {f_{B}}  {f_{BB}}  \,.
\label{eq:wilson_space_wbf}
\end{align}
With these basis vectors we calculate the Fisher information for
13~TeV using a combination of \toolfont{MadGraph5}~\cite{madgraph},
\toolfont{MadMax}~\cite{madmax2}, and our own \toolfont{MadFisher}
algorithm, described in Appendix~\ref{sec:algorithm}. We absorb
all particle identification and trigger efficiencies into a single universal $\varepsilon$
(which does not include the process-dependent CJV efficiencies). Then for our
toy example we assume the integrated luminosity times universal efficiencies to be 
$L \cdot \varepsilon  = 30~\ifb$. We find
\begin{align}
  I_{ij} (\mathbf{0}) =
\begin{pmatrix*}[r]
  3202.1 & -625.3 & -7.2 & -34.8 & 0.3 \\
  -625.3 & 451.0 & -109.5 & 23.3 & -1.5 \\
  -7.2 & -109.5 & 243.7 & -5.5 & 2.8 \\
  -34.8 & 23.3 & -5.5 & 4.1 & -0.3 \\
  0.3 & -1.5 & 2.8 & -0.3 & 0.1
\end{pmatrix*} \, .
\label{eq:information_wbf_tautau}
\end{align}
The eigenvectors, ordered by the size of their eigenvalues, are
\begin{align}
  \boldg_1 = \fivevec {0.98} {-0.21} {0.01} {-0.01} {0.00}  \quad 
  \boldg_2 = \fivevec {-0.18} {-0.79} {0.58} {-0.04} {0.01} \quad 
  \boldg_3 = \fivevec {0.12} {0.57} {0.81} {0.03} {0.01} \quad 
  \boldg_4 = \fivevec {0.00} {-0.05} {0.00} {1.00} {-0.07} \quad 
  \boldg_5 = \fivevec {0.00} {-0.00} {-0.01} {0.07} {1.00} \, .
\end{align}
The corresponding eigenvalues are $\left( 3338, 395, 165, 2.9, 0.1
\right)$, indicating that the WBF process has very different
sensitivities to the five operators: $\ope{\phi,2}$ can be most
strongly constrained and is weakly correlated with $\ope{W}$. It is
followed by the strongly correlated $\ope{W}$-$\ope{WW}$ plane.  The
sensitivity to $\ope{B}$ and $\ope{BB}$, which only play a role in
subleading $Z$-mediated production diagrams, is much smaller and shows
very little correlation with each other and everything else.\bigskip

We visualize our results as contours of the local and global distances
defined in Eq.\;\eqref{eq:distances} for slices of parameter space in
Fig.~\ref{fig:wbf_tautau_geometry}.  First, the contours show the
maximum precision that can be attained in a measurement in this
process. Without taking into account systematic uncertainties, an
optimal measurement will probe the $\ope{\phi,2}$ direction with
$\Delta g \approx 0.02$, translating into
$\Lambda/\sqrt{f_{\phi,2}} \approx 1.8$~TeV. The $\ope{W}$ and
$\ope{WW}$ directions can optimally be probed at the
$\Delta g \approx 0.05$ or $\Lambda/\sqrt{f_{\phi,2}} \approx 1.1$~TeV
level.

Comparing the local and global distances provides some insight into
the role of $\ord{1/\Lambda^4}$ effects, as discussed before.  At
$d = 1,2$ the differences are small, signaling that an optimal
measurement will be dominated by the linearized dimension-6
amplitudes. On the other hand, analyses based on less luminosity or
requiring more stringent exclusion criteria (translating into larger
distances) will only probe new physics scales closer to the
electroweak scale, in which case the squared dimension-6 terms will
have a larger effect.

\subsection{Differential information}

The fact that the Fisher information is additive across different
phase-space regions means that we can consider the differential
information with respect to phase space
($d I_{ij}/d\boldx$) or a specific kinematic
variable ($d I_{ij}/d\boldv$). In
Fig.~\ref{fig:wbf_tautau_differential_information} we show the
differential cross sections of the signal and dominant background
process for typical kinematic distributions and compare it to the
differential information.  More distributions are shown in
Appendix~\ref{sec:additional_plots}.

\begin{figure}
  \includegraphics[height=0.45 \textwidth]{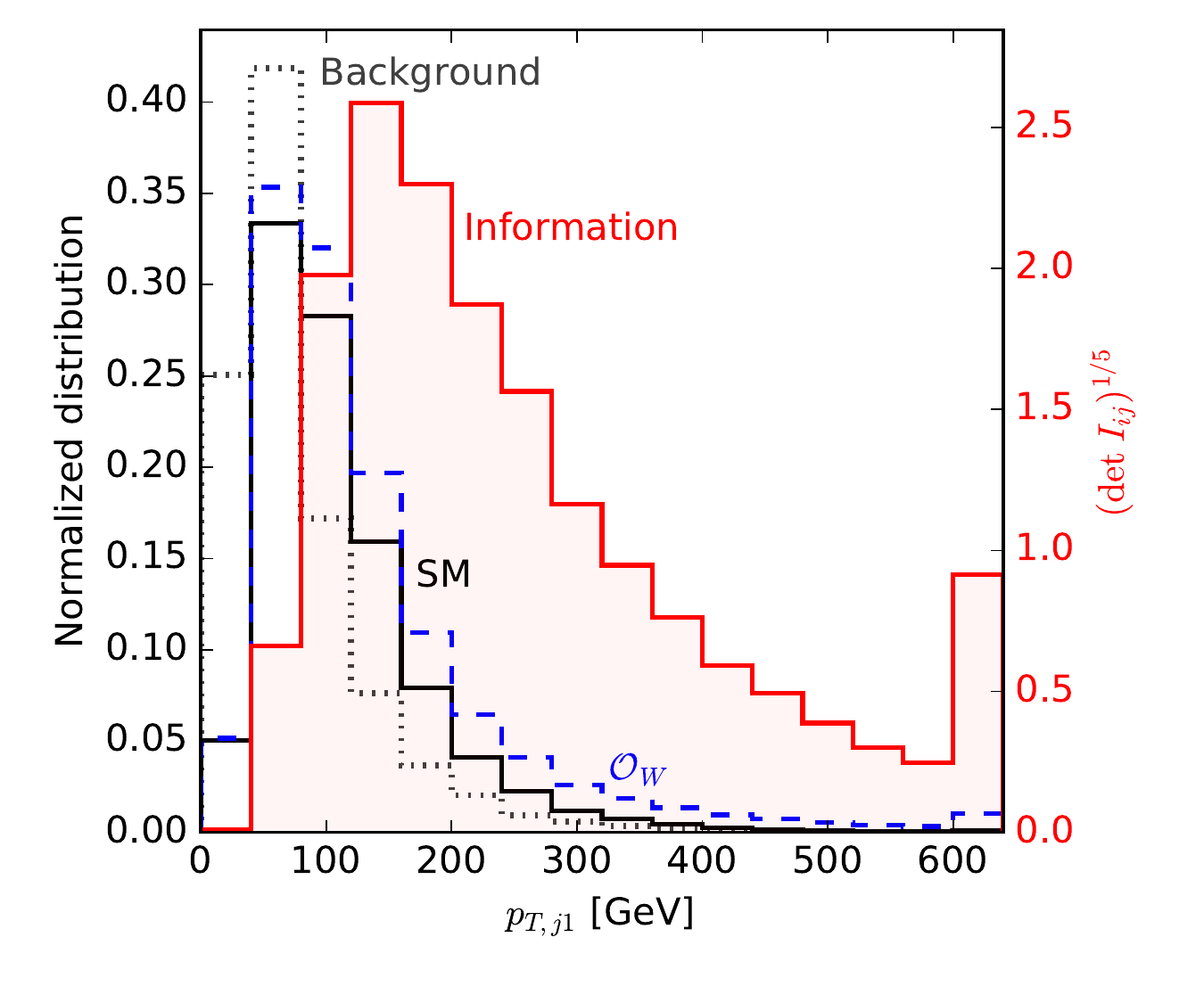}%
  \includegraphics[height=0.45 \textwidth]{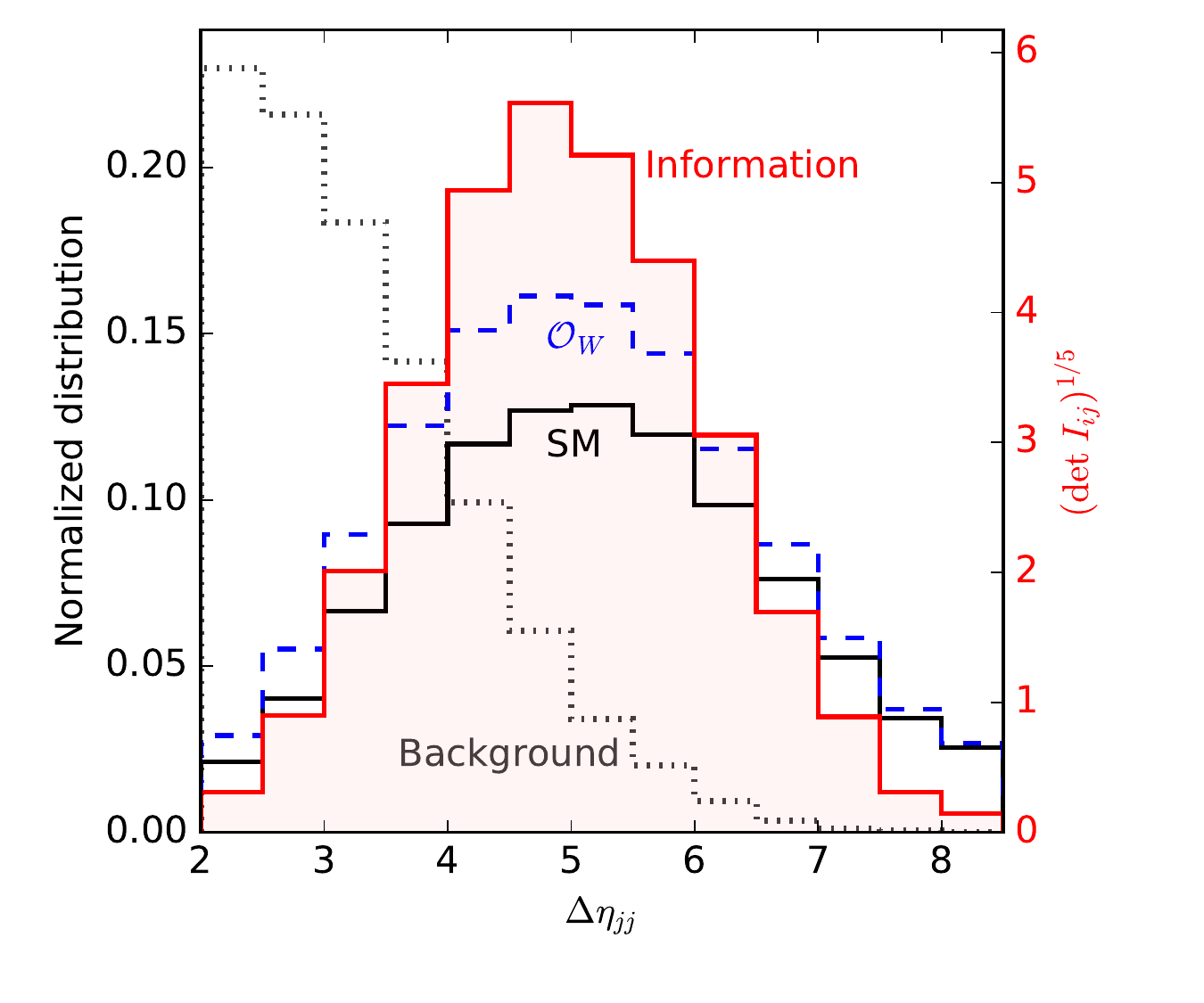}%
  \caption{Distribution of the Fisher information in the WBF $H \to
    \tau \tau$ channel (shaded red). We also show the normalized SM signal
    (solid black) and QCD $Z$+jets (dotted grey) rates. The dashed blue line
    shows the effect of an exaggerated $f_{W} \, v^2 / \Lambda^2 =
    0.5$. The last bin is an overflow bin.
    }
  \label{fig:wbf_tautau_differential_information}
\end{figure}


Obviously, the signal-to-background ratio improves for large invariant
masses of the tagging jets and towards $m_{\tau \tau}$ values around
the Higgs mass. The information is larger in these phase-space
regions, independent of the direction in model space.  On the other
hand, most of our dimension-6 operators include derivatives, leading
to an increasing amplitude with momentum transfer through the
gauge-Higgs vertex. This momentum flow is not observable, but the
transverse momenta of the tagging jets and the Higgs boson are
strongly correlated with it~\cite{eft-edge}. Indeed most of the
information on higher-dimensional operators comes from the high-energy
tail of $p_{T,j_1}$.

The rapidity difference between the tagging jets indicates a trade-off
between these two effects: on the one hand, at larger rapidity
distances the signal-to-background ratio clearly
improves~\cite{phi_jj}. On the other hand, the largest effects from
dimension-6 operators appear at smaller $\Delta \eta_{jj}$, again
driven by the larger momentum transfer~\cite{eft-edge}. In the right panel of
Fig.~\ref{fig:wbf_tautau_differential_information} we see that the
information on these operators comes from $\Delta \eta_{jj} =
3\dots7$. Tight cuts with the aim to remove backgrounds lose a sizable
fraction of the information on dimension-6 operators.

\subsection{Information in distributions}

\begin{figure}[b]
  \includegraphics[height=0.45 \textwidth]{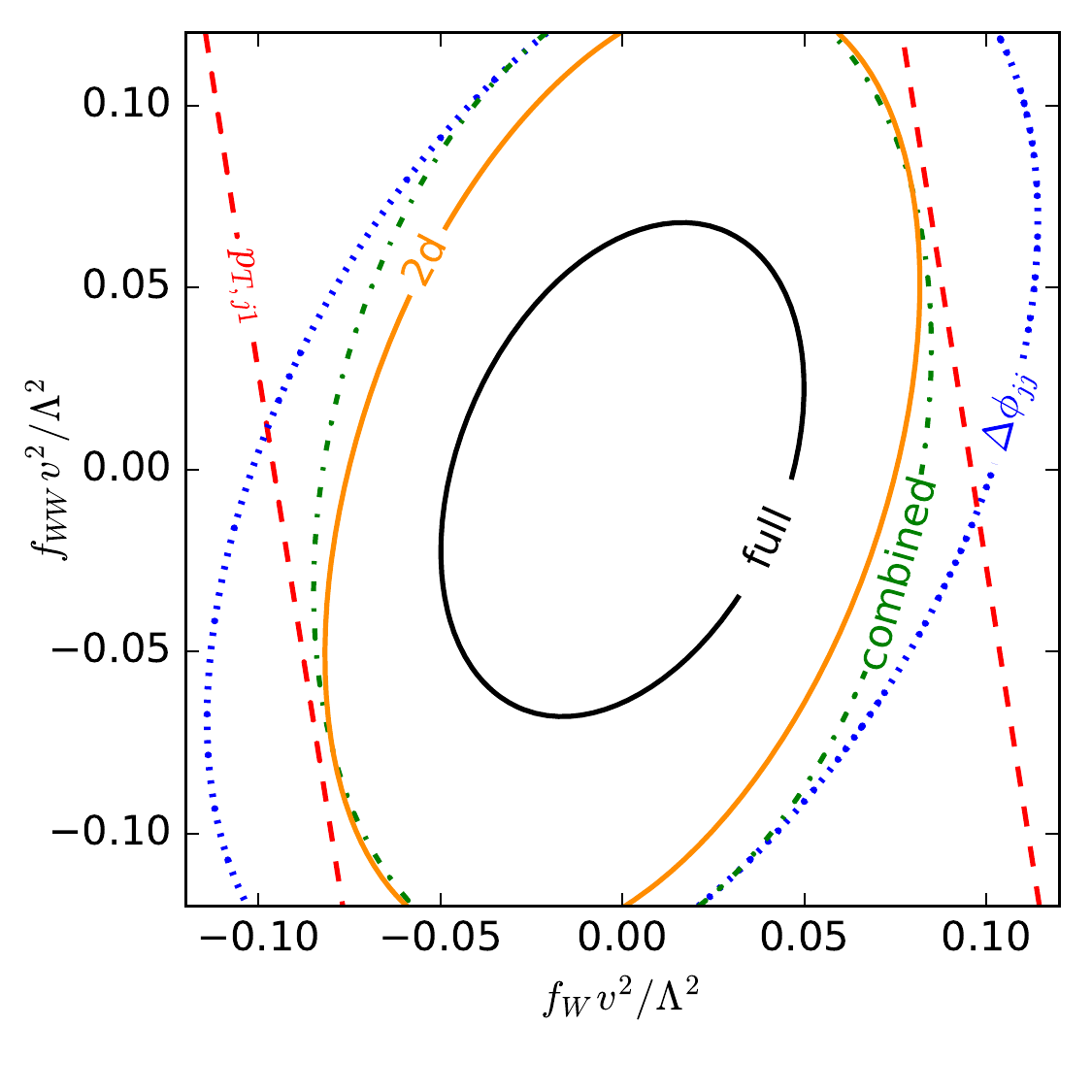}
  \caption{Information from histograms compared to the full
    information  (black) in the WBF $H \to \tau \tau$ channel, shown as contours
    $d_{\text{local}}(\boldg ; \boldzero) = 1$. We include
    $p_{T,j_1}$, $\Delta \phi_{jj}$, their naive combination assuming
    no mutual information, and their two-dimensional histogram. The
    $g_i$ not shown are set to zero.}
  \label{fig:wbf_tautau_histograms_contours}
\end{figure}

While the integrated, fully differential information defined in
Eq.\;\eqref{eq:fisher_rates} provides us with optimal experimental
results, it remains to be shown that we can access it in
practice. Recent proposals using machine learning for high-dimensional
likelihood fits aim to tackle exactly this problem~\cite{machine_learning}.
Regardless, a relevant question is how much of this maximum
information is retained in simple one-dimensional or two-dimensional
distributions of standard kinematic observables $\boldv$. 

In the presence of backgrounds, a histogram-based analysis first
requires a stringent event selection, either based on traditional
kinematic cuts or on a multivariate classifier. First, we choose the
WBF cuts
\begin{align}
  105~\gev < m_{\tau \tau} < 165~\gev \qquad
  p_{T,j_1} > 50~\gev \qquad
  m_{jj} > 1~\tev \qquad
  \Delta \eta_{jj} > 3.6 \,.
  \label{eq:wbf_tautau_wbfcuts}
\end{align}
This improves the signal-to-background ratio to approximately unity,
but at the cost of losing discrimination power. Eventually, a
histogram-based analysis will benefit from optimizing this selection,
for instance foregoing the simple cuts for a multivariate approach,
going beyond the scope of this demonstration.  Based on this
selection, we analyze the distributions
\begin{itemize}[label=\raisebox{0.1ex}{\scriptsize$\bullet$}]
\item $p_{T,\tau_1}$ with bin size 25~GeV up to 500~GeV and an
  overflow bin;
\item $m_{\tau \tau}$ with bin size 5~GeV in the allowed range of
  $105~...~165~\gev$;
\item $p_{T,\tau \tau}$ with bin size 50~GeV up to 800~GeV and an
  overflow bin;
\item $p_{T,j_1}$ with bin size 50~GeV up to 800~GeV and an
  overflow bin;
\item $m_{jj}$ with bin size 250~GeV up to 4~TeV and an overflow
  bin;
\item $\Delta \eta_{jj}$ with bin size $0.5$ up to $8.0$ and an
  overflow bin;
\item $\Delta \phi_{jj} = \phi_{j_{\eta < 0}} - \phi_{j_{\eta > 0}}$~\cite{phi_jjs} with bin size $2 \pi / 20$;
\item $\Delta \eta_{\tau\tau, j1}$ with bin size $0.5$ up to $8.0$ and an
  overflow bin;
\item $\Delta \phi_{\tau \tau, j1}$ with bin size $\pi / 10$.
\end{itemize} \bigskip 

\begin{figure}[t!]
  \includegraphics[height=0.6 \textwidth]{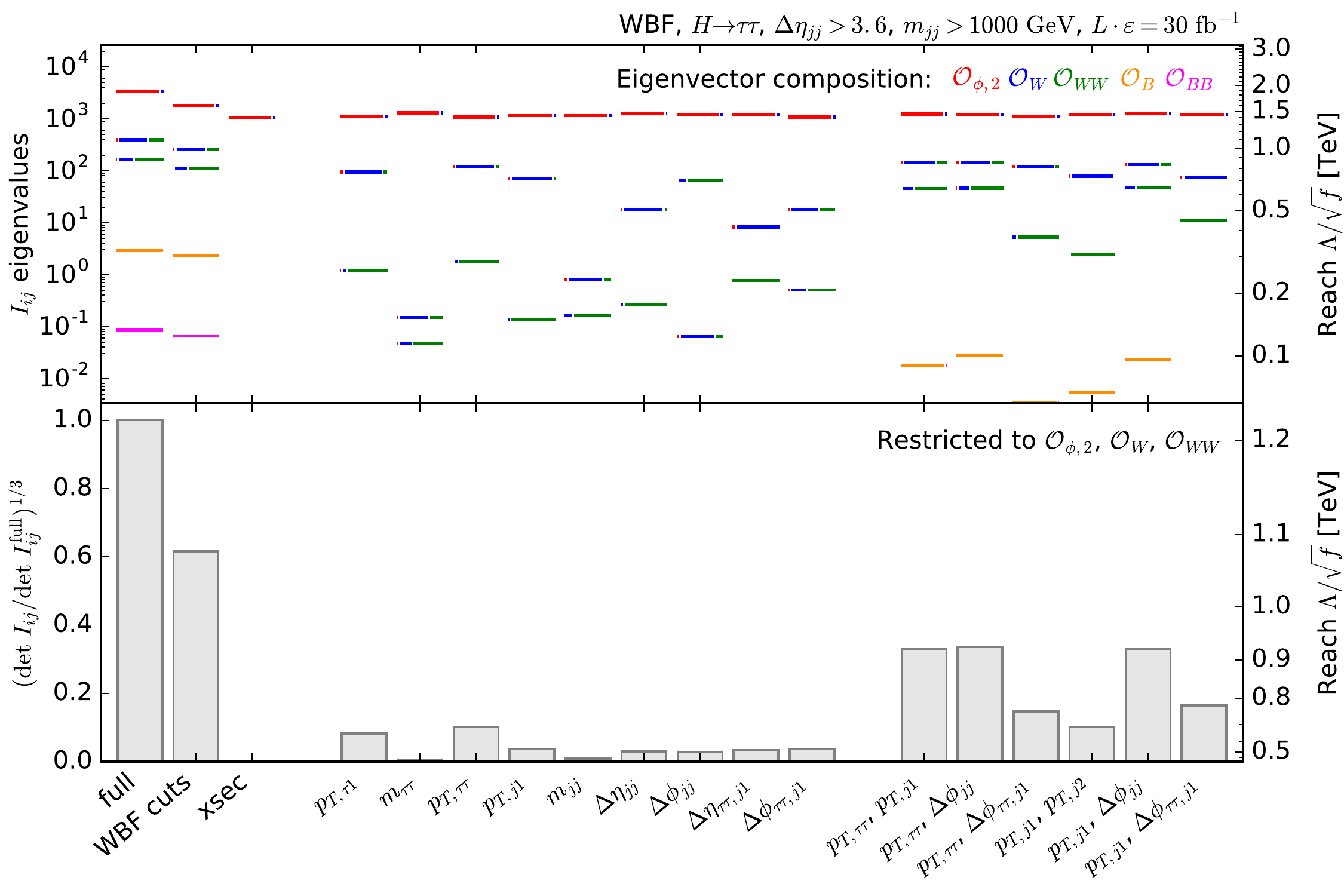}
  \caption{Fisher information for the WBF $H \to \tau \tau$ channel
    exploiting the full phase space, after the cuts in
    Eq.\;\eqref{eq:wbf_tautau_wbfcuts}, and for several kinematic
    distributions.  The top panel shows the eigenvalues, the colors
    denote the composition of the corresponding eigenvectors. The
    right axis translates the eigenvalues into a new physics reach for
    the corresponding combination of Wilson coefficients.  In the
    bottom panel we show the determinants of the Fisher information
    restricted to $\ope{\phi,2}$, $\ope{W}$, and $\ope{WW}$,
    normalized to the full information. Again, the right axis
    translates them into a new physics reach.}
\label{fig:wbf_tautau_histograms_comparison}
\end{figure}

Figure~\ref{fig:wbf_tautau_histograms_contours} demonstrates that
virtuality measures such as the transverse momentum of the leading
tagging jet mostly constrain $\ope{W}$, while angular correlations
between the jets are more sensitive to $\ope{WW}$. Stringent
constraints on the full operator space can only be achieved by
combining the information in these (or more) distributions, ideally in
a two-dimensional histogram.\bigskip

In Fig.~\ref{fig:wbf_tautau_histograms_comparison} we extend our
comparison to the information in all of the above distributions. The
top panel shows the eigenvalues of the individual information
matrices, and the colors indicate which operators the corresponding
eigenvectors are composed of. This allows us to see which operators
can be measured well in which distributions, and where blind (or flat)
directions arise. In the lower panel we compare the determinants,
providing a straightforward measure of the information in
distributions that is invariant under basis rotations. 

In general, single differential cross sections probe individual
directions in phase space well, but always suffer from basically blind
directions. To maximize the constraining power on all operators we
need to combine virtuality measures and angular correlations. Even
then there is a substantial difference to the maximum information in
the process: the combined analysis of jet transverse momenta and
$\Delta \phi_{jj}$ has a new physics reach in the
$\ope{\phi,2}$-$\ope{W}$-$\ope{WW}$ space of $0.9~\tev$, compared to
$1.2~\tev$ for the fully differential cross section.  Under our
simplistic assumptions this corresponds to roughly three times more
data.  Half of this loss in constraining power is due to information
in background-rich regions discarded by the WBF cuts, and half is due
to non-trivial kinematics not captured by the double differential
distributions. \bigskip

In light of the large amount of information discarded by the WBF cuts
in Eq.\;\eqref{eq:wbf_tautau_wbfcuts}, we repeat this comparison with
an alternative multivariate event selection. Instead of cutting on
standard kinematic observables, we select all events in
``signal-like'' phase-space regions, defined as those with a larger
expected SM WBF rate than expected background rates,
\begin{equation}
  \frac{\sigma_{\text{SM WBF}}\;f^{(1)}(\boldx|\text{SM WBF})}{\sigma_{\text{backgrounds}}\;f^{(1)} (\boldx|\text{backgrounds})}
  = \frac{\Delta \sigma_\text{SM WBF}(\boldx)}{\Delta \sigma_\text{backgrounds} (\boldx)}
  > 1 \,.
  \label{eq:wbf_tautau_likelihoodcuts}
\end{equation}
We then calculate the information in the same distributions as before.

\begin{figure}
  \includegraphics[height=0.6 \textwidth]{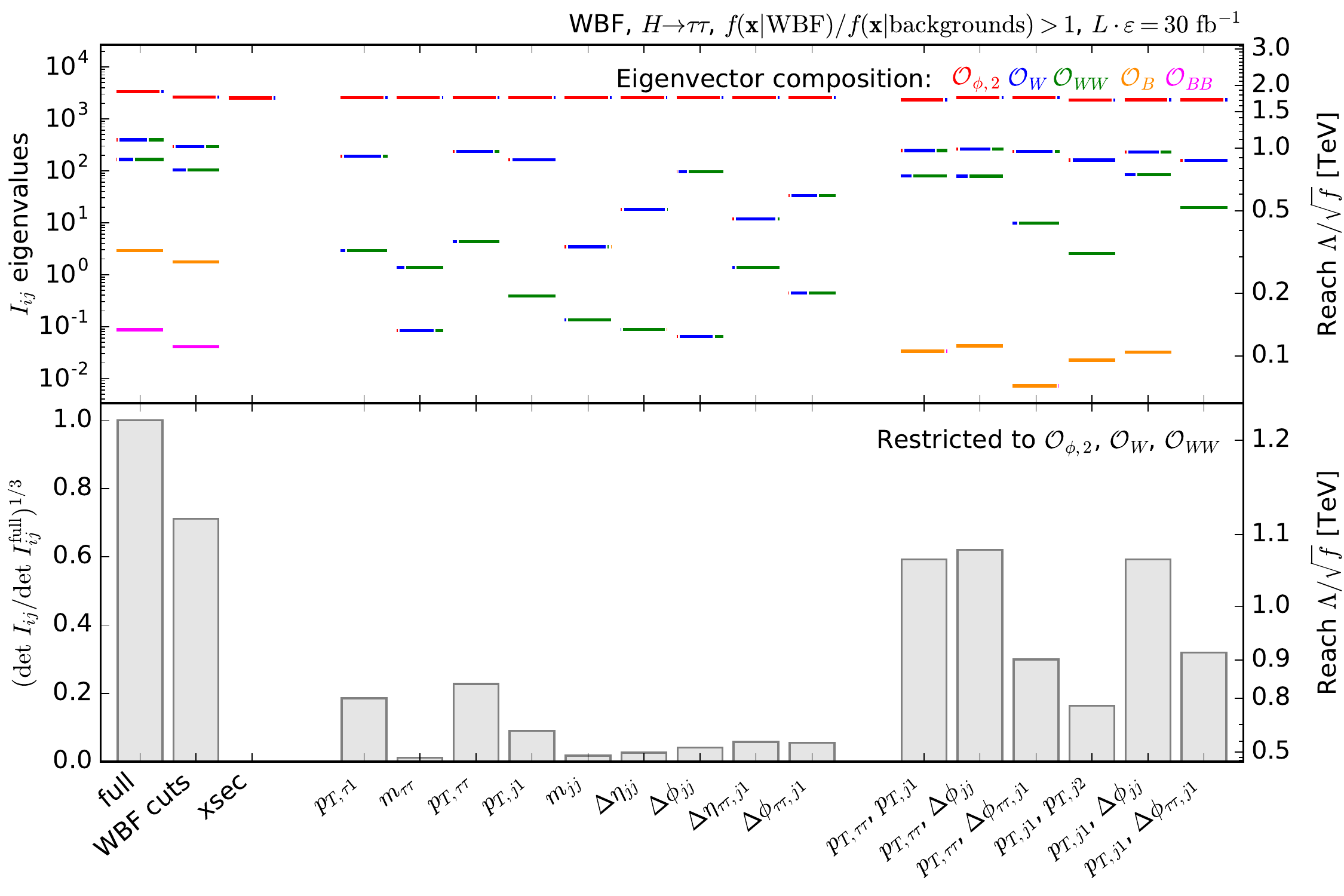}
  \caption{Fisher information for the WBF $H \to \tau \tau$ channel
    exploiting the full phase space, after the likelihood-based event
    selection in Eq.\;\eqref{eq:wbf_tautau_likelihoodcuts}, and for
    several kinematic distributions. Except for the initial cuts, the
    plot is analogous to
    Fig.~\ref{fig:wbf_tautau_histograms_comparison}.}
\label{fig:wbf_tautau_histograms_comparison_likelihoodcut}
\end{figure}

As shown in
Fig.~\ref{fig:wbf_tautau_histograms_comparison_likelihoodcut}, the cut
in Eq.\;\eqref{eq:wbf_tautau_likelihoodcuts} defines a sample with
little background contamination without sacrificing much
discrimination power. One-dimensional and two-dimensional
distributions can extract information on the operators more reliably
than after the kinematic event selection in
Eq.\;\eqref{eq:wbf_tautau_wbfcuts}. A combined measurement of the jet
transverse momenta and $\Delta \phi_{jj}$ is now able to probe new
physics scales of up to 1.1~GeV compared to 1.2~GeV for the fully
multivariate approach, corresponding to $70\%$ more data.

\section{Weak-boson-fusion Higgs to four leptons}
\label{sec:wbf_4l}

Another question we can approach with information geometry is how much
the non-trivial decay mode $H \to 4 \ell$ adds to the WBF production
analyzed in Sec.~\ref{sec:wbf_taus}. For this particularly clean
channel, shown in Fig.~\ref{fig:wbf_4l_diag}, the backgrounds are not
the limiting factor, so we omit them for our toy study. For instance,
in the relevant phase-space region the cross section of the dominant
irreducible $ZZ^* \,jj$ background is over an order of magnitude
smaller than the SM Higgs signal.  This also allows us to avoid
smearing the $m_{4\ell}$ distribution. At parton level we apply the
generator-level cuts
\begin{alignat}{2}
  p_{T,j} &> 20 \ \gev \qquad & \qquad |\eta_{j}| &< 5.0  \notag \\ 
  p_{T,\ell} &> 10 \ \gev  \qquad & \qquad |\eta_{\ell}| &< 2.5 \; ,
\label{eq:wbf_4l_acceptance_cuts}
\end{alignat}
with $\ell = e, \mu$. The SM cross section after these cuts is 0.36~fb. 

\begin{figure}[b]
  \includegraphics[width=0.44 \textwidth]{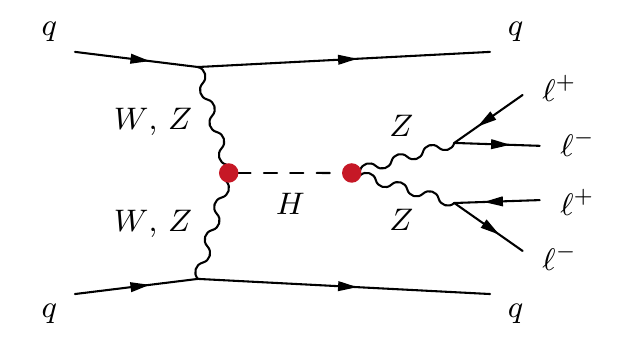}%
  \caption{Example Feynman diagram for weak-boson-fusion Higgs production with
    $H \to 4 \ell$. The red dots show the Higgs-gauge interactions
    affected by the dimension-6 operators of our analysis.}
\label{fig:wbf_4l_diag}
\end{figure}

\subsection{Maximum precision on Wilson coefficients}

Again we study the five-dimensional space of CP-even Wilson
coefficients given in Eq.\;\eqref{eq:wilson_space_wbf}.  For increased
luminosity, $L \cdot \varepsilon = 100~\ifb$, we find the SM
information
\begin{align}
  I_{ij} (\boldzero) =
\begin{pmatrix*}[r]
  144.3 & -27.3 & -11.5 & -1.6 & -0.7 \\
  -27.3 & 50.9 & -9.1 & 6.7 & -0.2 \\
  -11.5 & -9.1 & 36.9 & -1.2 & 1.0 \\
  -1.6 & 6.7 & -1.2 & 1.9 & -0.1 \\
  -0.7 & -0.2 & 1.0 & -0.1 & 0.1
\end{pmatrix*}
\end{align}
with the eigenvectors 
\begin{align}
  \boldg_1 = \fivevec {0.96} {-0.25} {-0.08} {-0.02} {0.00}  \quad 
  \boldg_2 = \fivevec {-0.16} {-0.79} {0.58} {-0.11} {0.02}  \quad
  \boldg_3 = \fivevec {0.21} {0.54} {0.81} {0.09} {0.02} \quad 
  \boldg_4 = \fivevec {0.02} {0.14} {0.01} {-0.99} {0.04}  \quad 
  \boldg_5 = \fivevec{0.00} {-0.00} {-0.03} {0.04} {1.00}  \,.
\end{align}
and the eigenvalues $\left( 152.4, 52.8, 27.8, 1.0, 0.0 \right)$. 

The Fisher information approach allows us to directly compare this
outcome to our earlier results for WBF production with
$H \to \tau \tau$ in Eq.~\eqref{eq:information_wbf_tautau}, or to
calculate the combined information in these two channels by simply
adding their Fisher information matrices after rescaling them to the
same luminosity. Clearly, the $\tau \tau$ channel contains
significantly more information on all operators. The decay
$H \to 4\ell$ does not even increase the sensitivity to $\ope{B}$ or
$\ope{BB}$, both of them are still basically blind directions. We
visualize the information geometry in the remaining directions in
Fig.~\ref{fig:wbf_4l_geometry}. The differences between local and
global distances are much larger than in the $H \to \tau \tau$
channel. This is because the tiny $H \to 4\ell$ branching fraction
decreases the new physics reach and with it the hierarchy of scales in
our effective Lagrangian. This means that the squared dimension-6
amplitudes are numerically more relevant.

\begin{figure}
  \includegraphics[height=0.34 \textwidth,clip,trim=0.3cm 0 0.05cm 0]{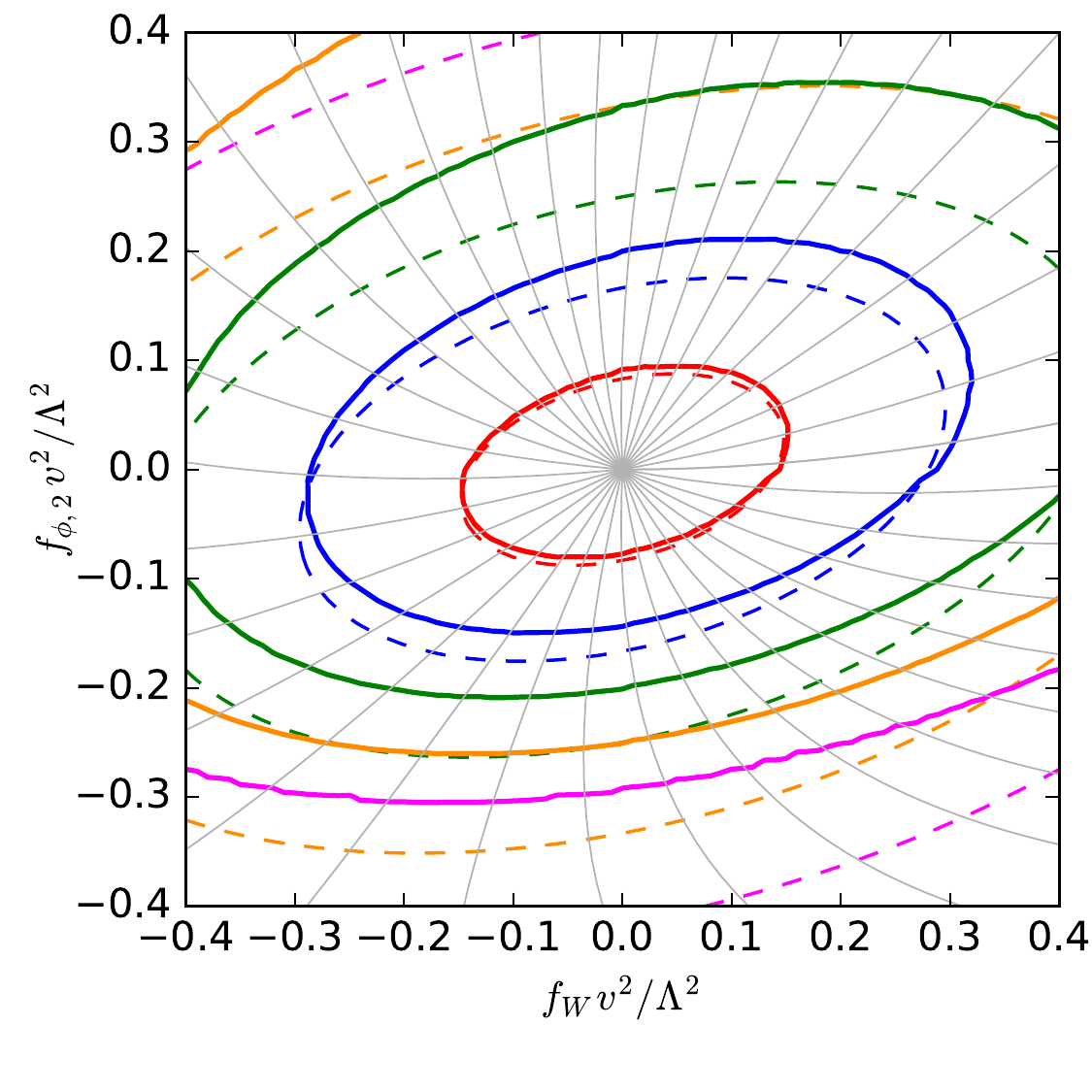}%
  \includegraphics[height=0.34 \textwidth,clip,trim=0.3cm 0 0.05cm 0]{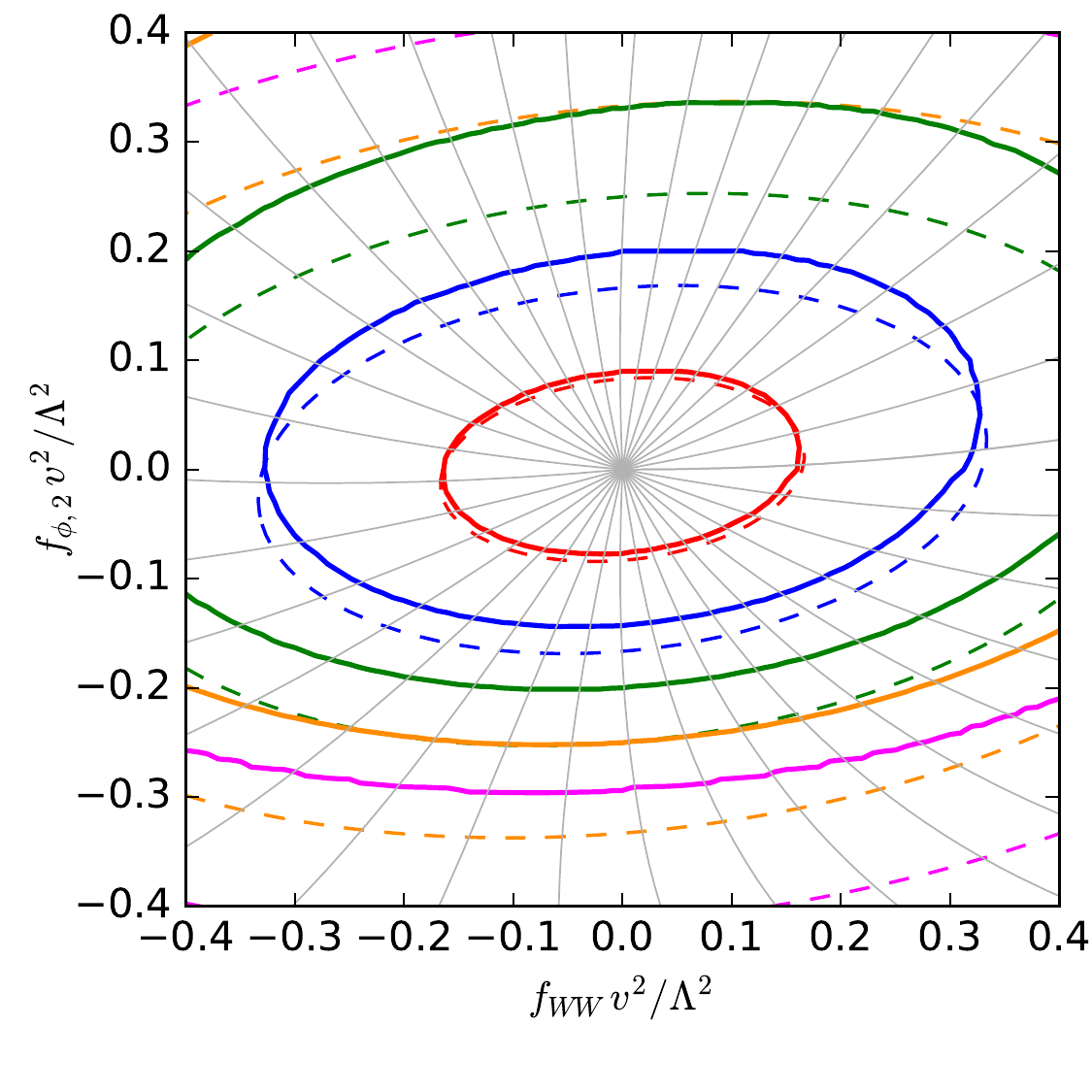}%
  \includegraphics[height=0.34 \textwidth,clip,trim=0.3cm 0 0.05cm 0]{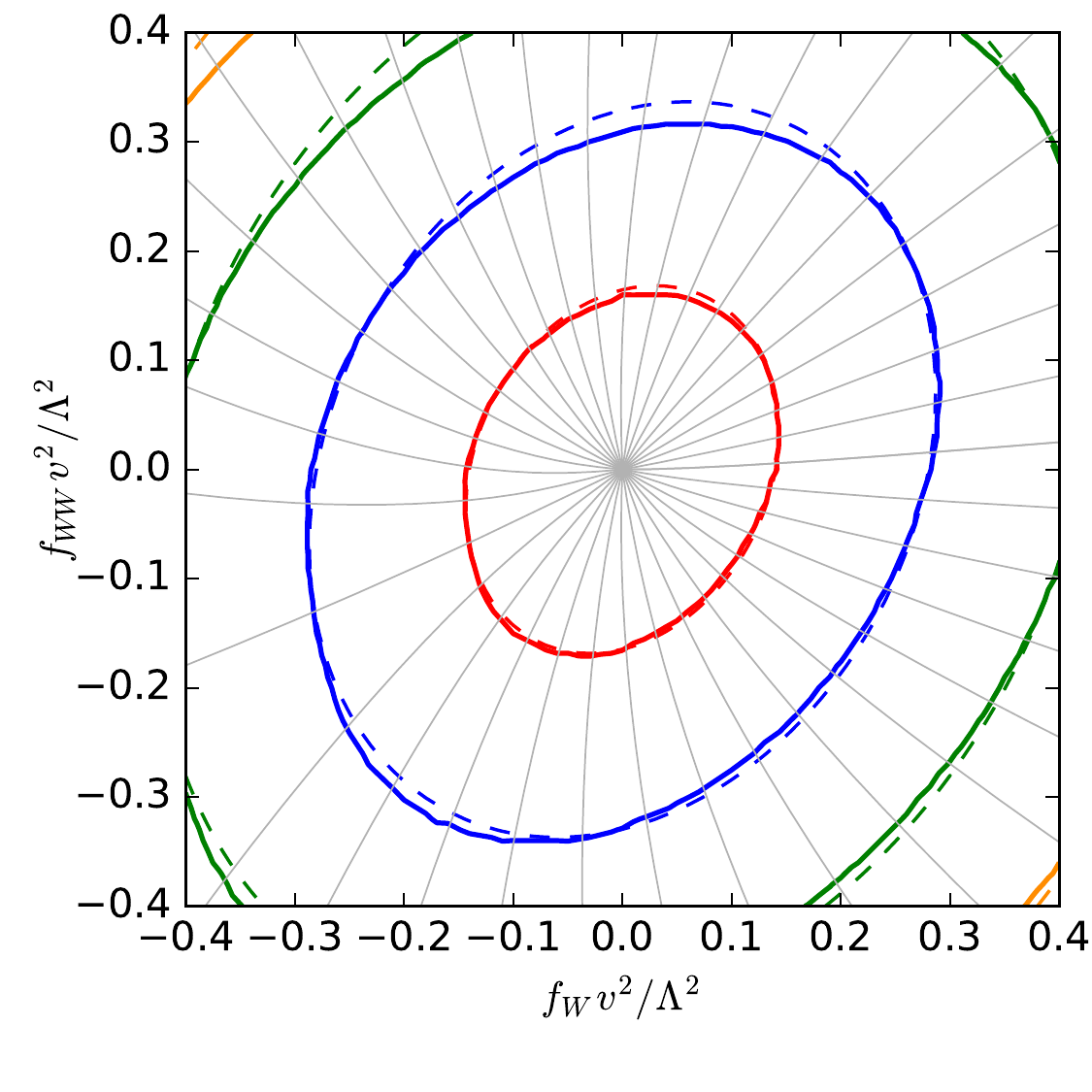}%
  \caption{Error ellipses defined by the Fisher information in the
    WBF $H \to 4\ell$ channel. We show contours of local distance
    $d_\text{local}(\boldg ; \boldzero)$ (dashed) and global
    distance $d(\boldg,\boldzero)$ (solid). The colored contours
    indicate distances of $d = 1~...~5$. In grey we show example
    geodesics.  The $g_i$ not shown are set to zero.}
\label{fig:wbf_4l_geometry}
\end{figure}

\subsection{Production vs decay kinematics}

\begin{figure}[b]
  \includegraphics[height=0.45 \textwidth]{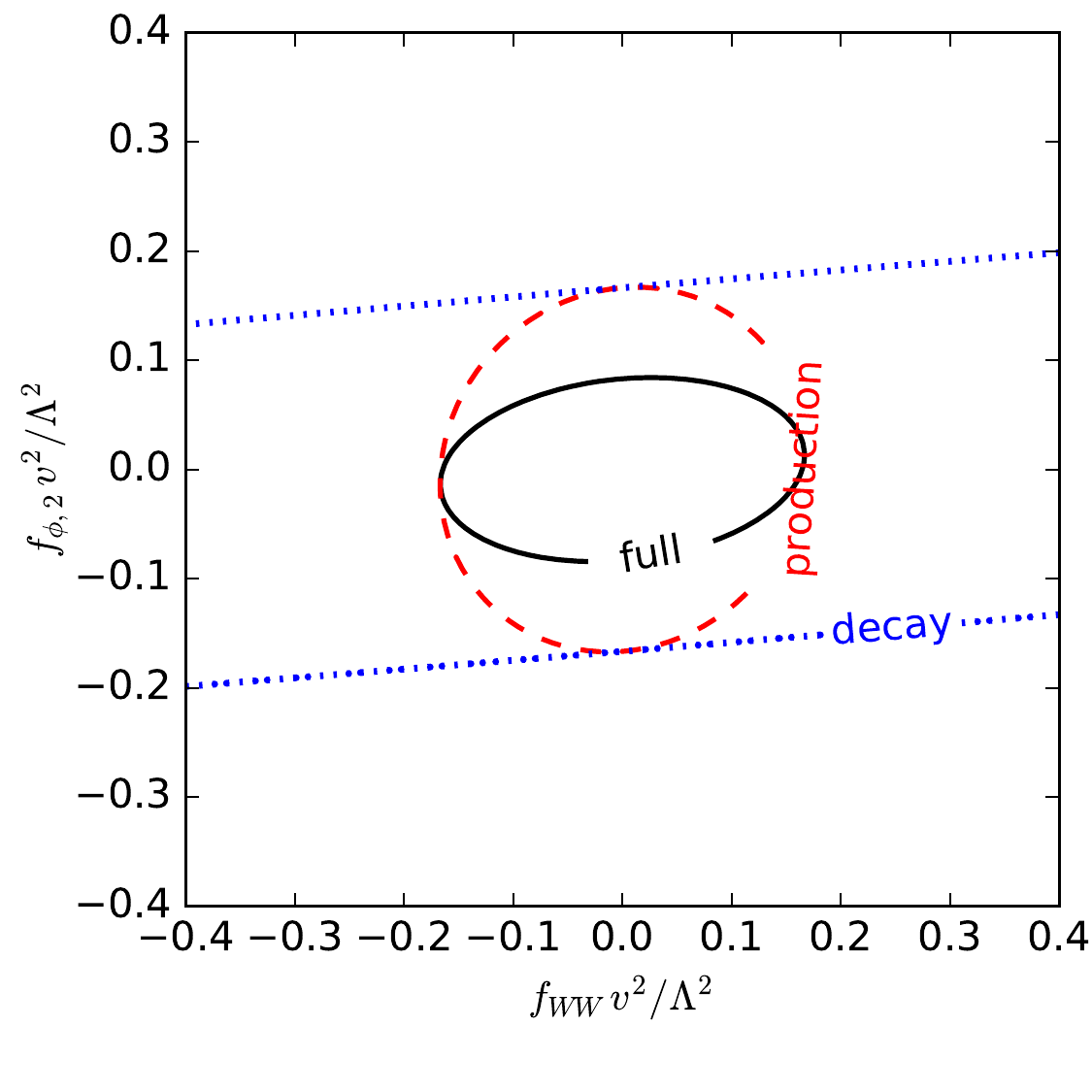}
  \caption{Information in the WBF $H \to 4\ell$ channel from including
    dimension-6 operators only in the production vertex (red), only in
    the decay vertex (blue), and both (black). The information is
    visualized as local contours
    $d_{\text{local}}(\boldg ; \boldzero) = 1$. The $g_i$ not shown
    are set to zero.}
\label{fig:wbf_4l_production_decay}
\end{figure}

Focusing on the question how the decay analysis improves our global
information, we disentangle the effects on the production and decay
vertices in Fig.~\ref{fig:wbf_4l_production_decay}.  As well known for
the LHC, the production-side analysis benefits from a large momentum
flow through the Higgs vertex, while the momentum flow through the
decay vertices is bounded by the Higgs mass (neglecting off-shell
phase space regions).  For momentum-dependent operators this
disadvantage is not compensated by the complex $H \to 4\ell$ decay
kinematics.  Consequently, the Higgs decay only improves the reach in
the $\ope{\phi,2}$ direction, corresponding to a change in the total rate. 
This operator also affects many other total Higgs rates,
so we conclude that the complex $H \to 4\ell$ kinematics
does not play a significant role as part of a global analysis.\bigskip 

\begin{figure}[t!]
  \includegraphics[height=0.6 \textwidth]{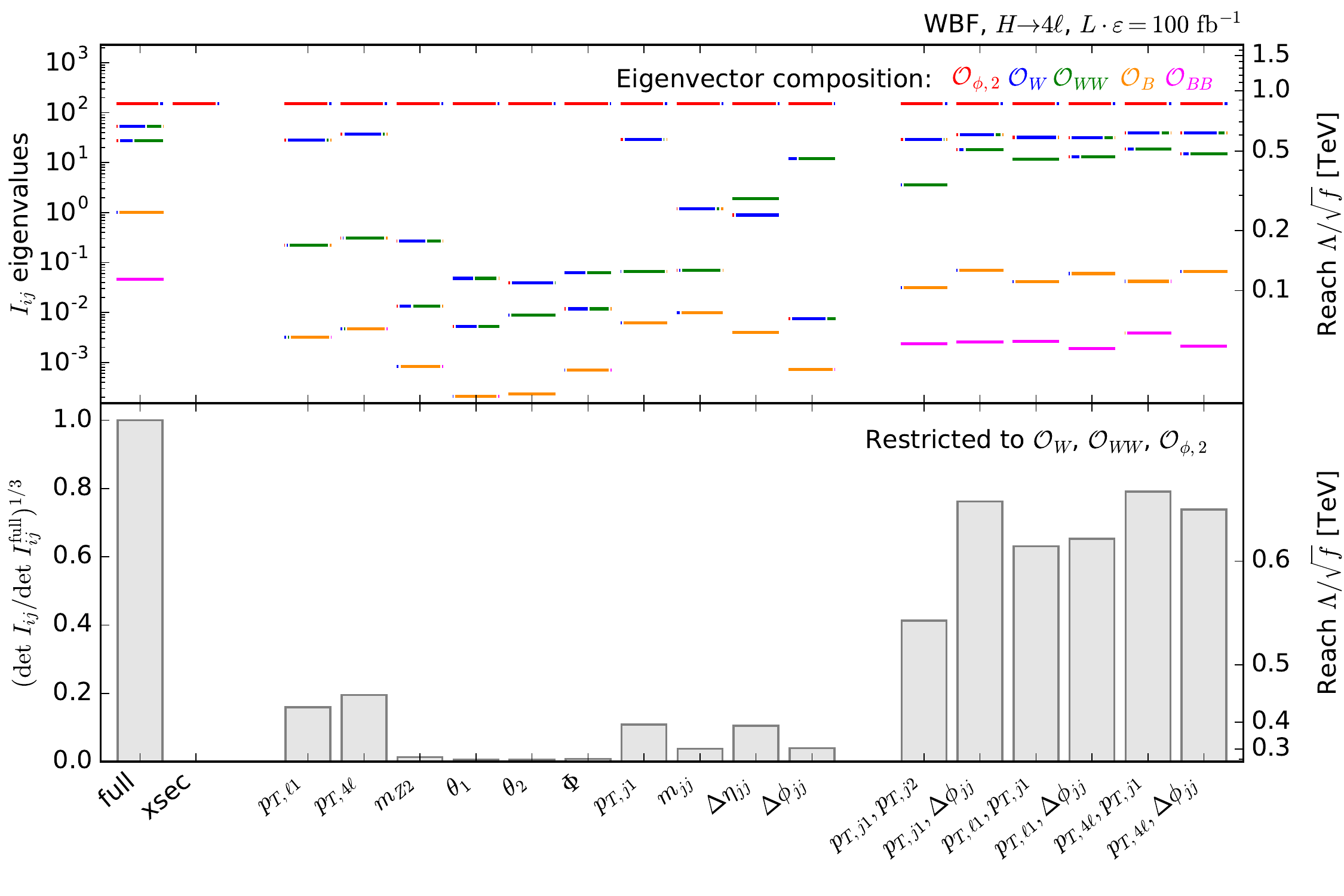}
  \caption{Fisher information for the WBF $H \to 4 \ell$ channel,
    based on the full kinematics and on individual kinematic
    distributions. The top panel shows the eigenvalues, the colors
    denote the composition of the corresponding eigenvectors. The
    right axis translates the eigenvalues into a new physics reach for
    the corresponding combination of Wilson coefficients.  In the
    bottom panel we show the determinants of the Fisher information
    restricted to $\ope{\phi,2}$, $\ope{W}$, and $\ope{WW}$,
    normalized to the full information. Again, the right axis
    translates them into a new physics reach.}
\label{fig:wbf_4l_histograms_comparison}
\end{figure}

In complete analogy to Fig.~\ref{fig:wbf_tautau_histograms_comparison}
for the WBF production, we compare the information in different
distributions for $CP$-even operators in
Fig.~\ref{fig:wbf_4l_histograms_comparison}.  The standard tagging jet
observables are complemented by five observables characterizing the
$4\ell$ decay kinematics,
\begin{itemize}[label=\raisebox{0.1ex}{\scriptsize$\bullet$}]
  \item $p_{T,\ell_1}$;
  \item $p_{T,4\ell}$;
  \item $m_{Z_2}$ for the lower-mass reconstructed $Z$ boson;
  \item $\cos \theta_1 = \hat{p}_{\ell^-_1} \cdot \hat{p}_{Z_2}
    \Big|_{Z_1}$ defined in terms of unit-3-vectors $\hat{p}$, and analogously
    $\cos \theta_2$;
  \item $\cos \Phi = ( \hat{p}_{\ell^-_1} \times
    \hat{p}_{\ell^+_1} ) \cdot ( \hat{p}_{\ell^-_2} \times
    \hat{p}_{\ell^+_2} )$, defined in the $ZZ$ or Higgs rest
    frame~\cite{phi_jj}.
\end{itemize}
In all cases we use at least ten bins and include underflow and
overflow bins where applicable.

In our quantitative analysis we find similar patterns as in the
$\tau \tau$ mode. The key observables are again transverse momenta and
jet angular correlations. Without the complication of removing
backgrounds efficiently, the combined analysis of these variables
comes close to the maximum information: a two-dimensional histogram of
jet transverse momenta and $\Delta \phi_{jj}$ probes new physics
scales up to 650~GeV, while for a fully differential analysis the
maximum probed new physics scale is close to 700~GeV. Under our
assumptions, this difference roughly corresponds to 25\% more
data. The decay kinematics and its angular observables do not help
significantly or change the picture qualitatively.  This shows again
how much the sensitivity of the decay vertices to dimension-6
operators is limited by the restriction of the momentum flow to the
Higgs mass. This is not accidental: the reason behind this role of the
momentum dependence is that for all operators shown in
Eq.\;\eqref{eq:wilson_space_wbf} with the exception of $\ope{\phi,2}$,
gauge invariance forces us to include the field strength tensor
instead of the gauge boson field, automatically introducing a momentum
dependence.

\section{Higgs plus single top}
\label{sec:th}

Our final example is Higgs production with a single top with
$H\to \gamma \gamma$ and a hadronic top decay. As shown in
Fig.~\ref{fig:th_diag}, diagrams where the Higgs is radiated off a $W$
boson interfere destructively with diagrams with a top-Higgs coupling,
making this channel a direct probe of the sign of the top Yukawa
coupling~\cite{top_higgs}. We stick to a parton-level analysis at
leading order in the five-flavor scheme. For our toy example we
include only one of the dominant backgrounds, single top production
with two photons, and in particular ignore the multi-jet
background. The subleading $t\bar{t} \, \gamma\gamma$ background
populates qualitatively different phase-space regions from the
single-top signal and can be supressed with an appropriate event
selection~\cite{Kling:2012up}. We smear the $m_{\gamma \gamma}$
distribution of the signal process with a Gaussian of width 1.52 GeV
estimated from Fig.~6b of Ref.~\cite{CMS:2016zjv}, and do not include
any other detector effects. Our basic event selection requires
\begin{alignat}{4}
  p_{T,j} &> 20 \ \gev \quad & \quad
  |\eta_{j}| &< 5.0 \quad & \quad
  \Delta R_{jj} &> 0.4 \quad & \quad
  152 \ \gev &<m_{bjj} < 192 \ \gev \notag \\ 
  p_{T,\gamma} &> 10 \ \gev \quad & \quad
  |\eta_{\gamma}| &< 2.5 \quad & \quad 
  \Delta R_{\gamma j} , \Delta R_{\gamma \gamma} &> 0.4 \quad & \quad
  120 \ \gev &<m_{\gamma \gamma} < 130 \ \gev \,,
  \label{eq:th_acceptance_cuts}
\end{alignat}
leading to a SM $tH$ cross section of 0.10~fb and a background of 0.22~fb.

\begin{figure}
  \includegraphics[width=0.773 \textwidth]{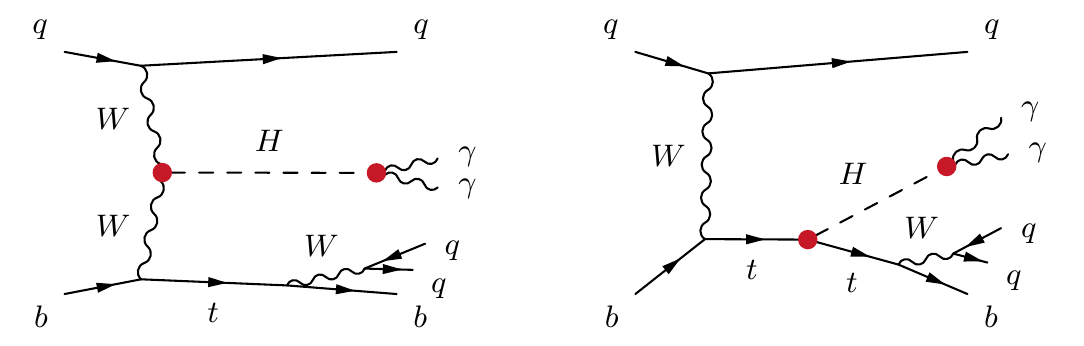}%
  \caption{Example Feynman diagrams for Higgs production with a single
    top, not resolving the loop-induced $H\gamma \gamma$ coupling. The red
    dots show the Higgs interactions affected by the dimension-6
    operators of our analysis.}
\label{fig:th_diag}
\end{figure}

We consider four $CP$-even dimension-6 operators
\begin{alignat}{2}
  \ope{W}  &= i \frac{g}{2} \, (D^\mu\phi)^\dagger \sigma^k ( D^\nu\phi) \, W_{\mu\nu}^k  \qquad & \qquad
  \ope{t \phi}  &= (\phisq) \, ( \overbar{Q}_3 \tilde{\phi} t_R)  + \text{h.c.} \notag \\
  \ope{WW}  &= -\frac{g^2}{4} \, (\phisq) \, W^k_{\mu\nu} \, W^{\mu\nu\, k}  \qquad & \qquad
  \ope{\phi,2}  &= \frac{1}{2} \, \partial^\mu(\phi^\dagger\phi) \, \partial_\mu(\phi^\dagger\phi) \,.
\label{eq:th_ope}
\end{alignat}
The operators $\ope{W}$ and $\ope{WW}$ affect the production amplitudes where
the Higgs couples to a $W$, while $\ope{t \phi}$ re-scales the top
Yukawa coupling. Both, $\ope{WW}$ and $\ope{\phi,2}$ also affect the
$H \to \gamma \gamma$ decay.

\subsection{Maximum precision on Wilson coefficients}

\begin{figure}[b]
  \includegraphics[height=0.34 \textwidth,clip,trim=0.3cm 0 0.05cm 0]{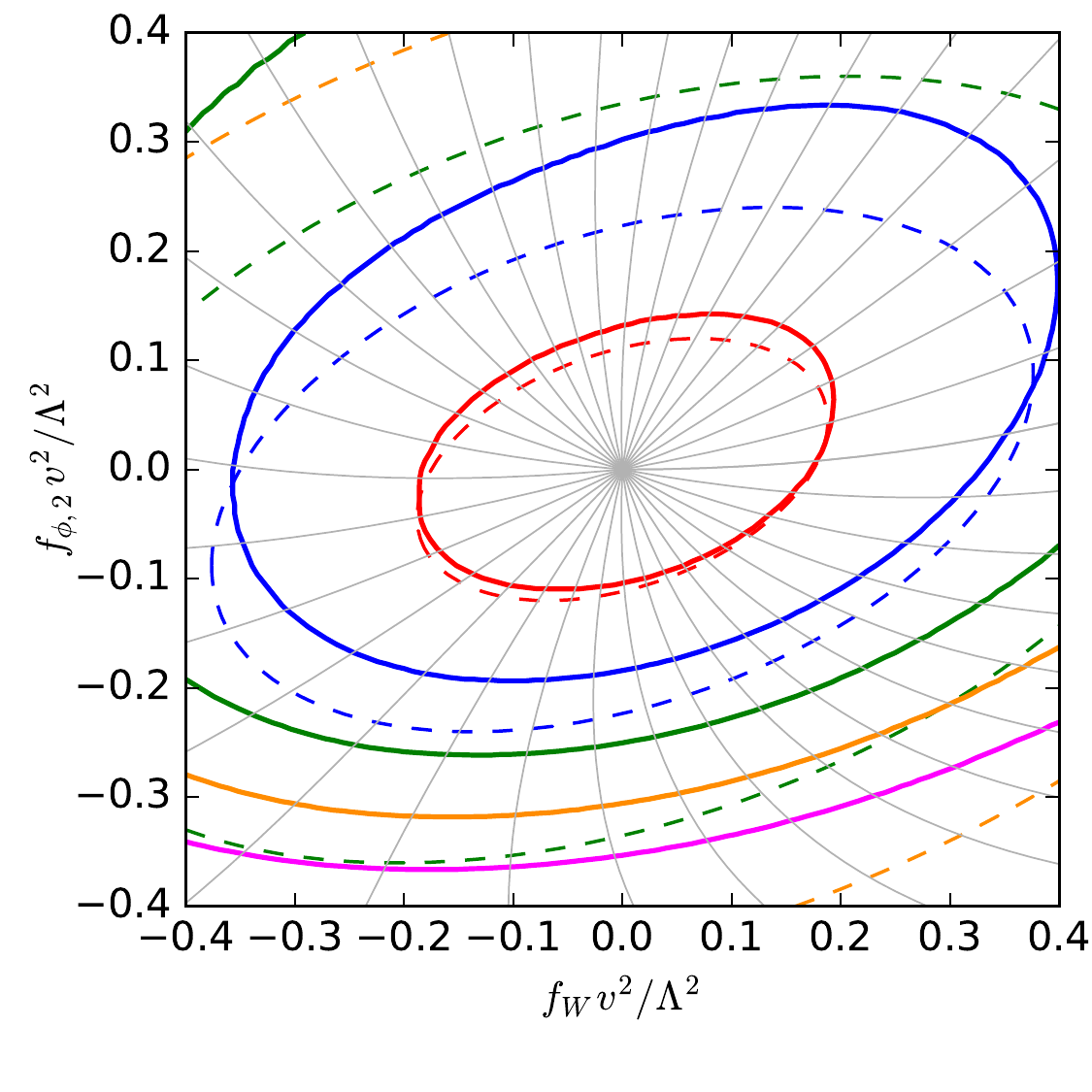}%
  \includegraphics[height=0.34 \textwidth,clip,trim=0.3cm 0 0.05cm 0]{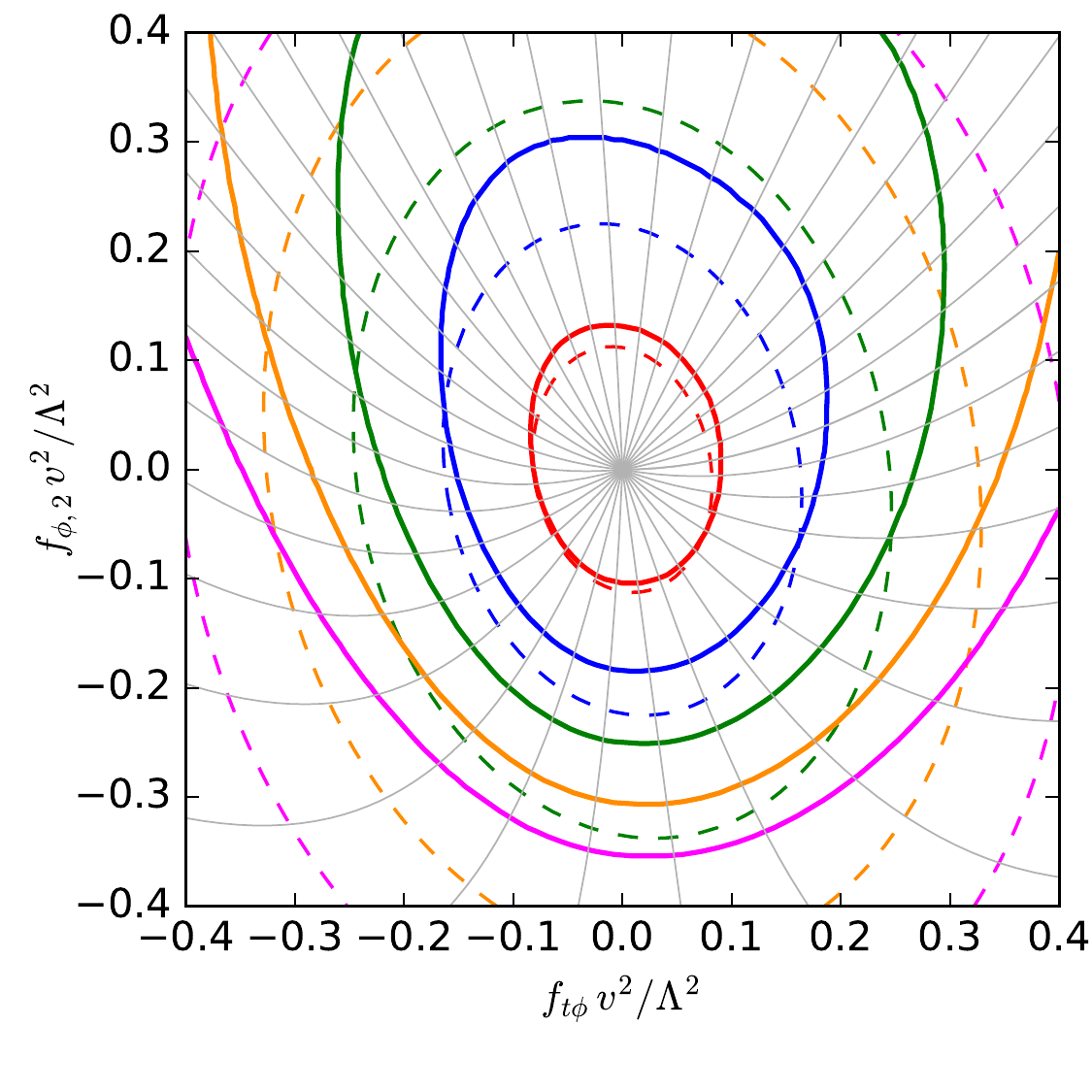}%
  \includegraphics[height=0.34 \textwidth,clip,trim=0.3cm 0 0.05cm 0]{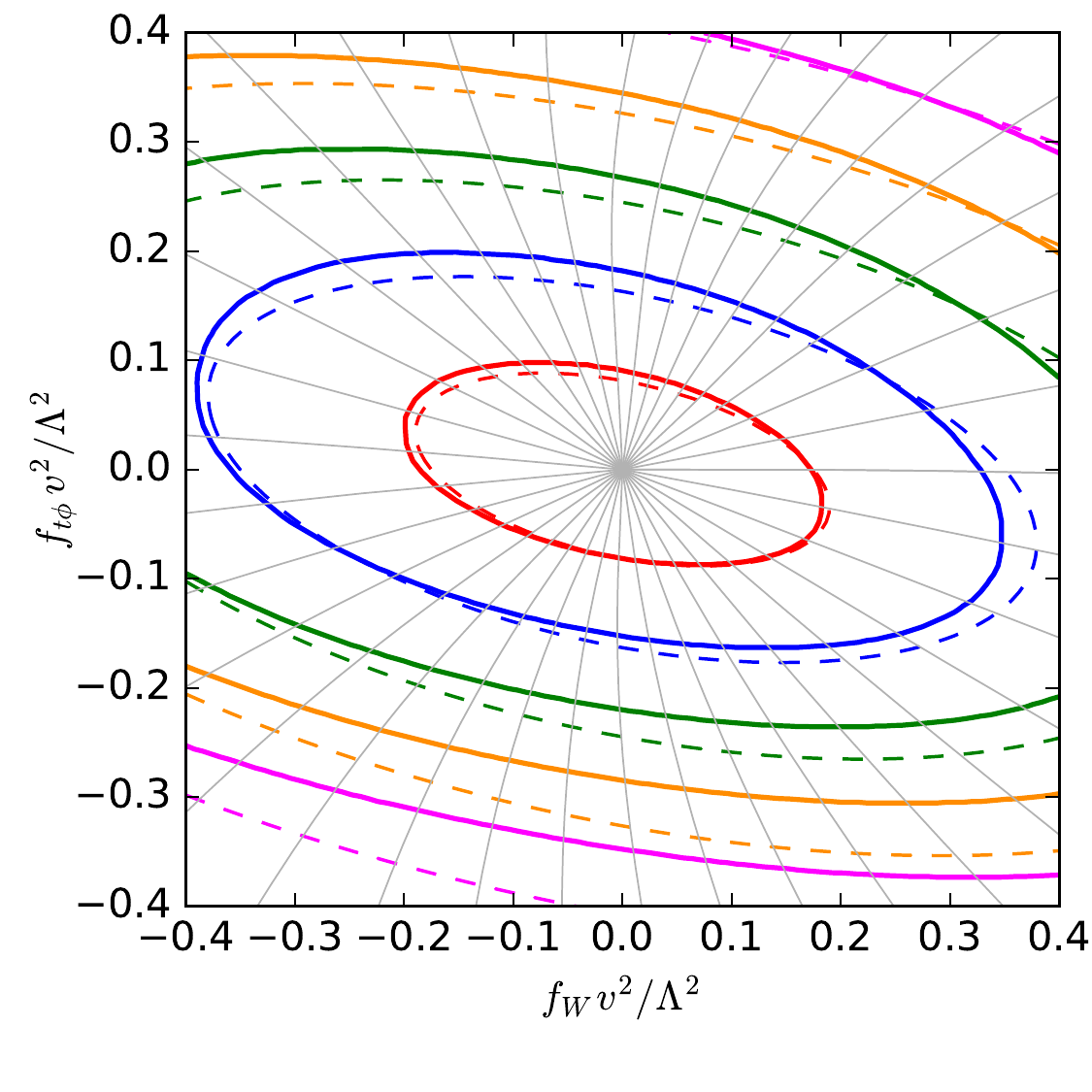}%
  \caption{Error ellipses defined by the Fisher information in Higgs
    plus single top production. We show contours of local distance
    $d_\text{local}(\boldg ; \boldzero)$ (dashed) and global distance
    $d(\boldg,\boldzero)$ (solid).  The colored contours indicate
    distances of $d = 1~...~5$. In grey we show example geodesics. The
    $g_i$ not shown are set to zero. }
\label{fig:th_geometry}
\end{figure}

We calculate the Fisher information in terms of the dimensionless parameters
\begin{align}
  \boldg = \frac {v^2} {\Lambda^2}  \fourvecc {f_{\phi,2}} {f_W} {f_{WW}} {f_{t \phi}}
  \label{eq:wilson_space_th}
\end{align}
for 13~TeV and an integrated luminosity times efficiencies of
$L \cdot \varepsilon = 300~\ifb$ and find
\begin{align}
  I_{ij} (\mathbf{0}) =
\begin{pmatrix*}[r]
  80.1 & -18.7 & -957.0 & 13.2 \\
  -18.7 & 32.6 & 221.7 & 27.0 \\
  -957.0 & 221.7 & 11446.1 & -146.0 \\
  13.2 & 27.0 & -146.0 & 150.3
\end{pmatrix*} \, .
\end{align}
The eigenvectors are
\begin{align}
  \boldg_1 = \fourvec {0.08} {-0.02} {-1.00} {0.01}  \qquad 
  \boldg_2 = \fourvec {0.00} {-0.23} {-0.01} {-0.97} \qquad 
  \boldg_3 = \fourvec {-0.02} {0.97} {-0.02} {-0.23} \qquad 
  \boldg_4 = \fourvec {1.00} {0.02} {0.08} {-0.01}
\end{align}
with corresponding eigenvalues $\left( 11532, 155, 21.3, 0.1 \right)$.
The best constrained direction is along $\ope{WW}$ and corresponds to
the combination of Wilson coefficients that affects the
$H \to \gamma \gamma$ decay in addition to production effects, which
will already be tightly constrained once a $tH$ measurement is
feasible. The orthogonal direction in the $\ope{\phi,2}$-$\ope{WW}$
plane is for all practical purposes blind. Even with the assumed
sizeable event rate corresponding to 300~$\ifb$, the sensitivity to
$\ope{W}$ and $\ope{t \phi}$ is limited, with some mixing between the
two operators.\bigskip

We visualize this maximum sensitivity to dimension-6 operators in
Fig.~\ref{fig:th_geometry}. With the exception of $\ope{WW}$, an
optimal measurement can probe all operators at the
$\Delta g \approx 0.1\dots 0.2$ level, equivalent to
$\Lambda/\sqrt{f_{\phi,2}} \approx 600 \dots 750$~GeV.  There are
large differences between local and global distances already at the
$d = 2$ level, implying that a measurement of this channel will always
be sensitive to the squared dimension-6 terms.

\subsection{Differential information}

\begin{figure}
  \includegraphics[height=0.45 \textwidth]{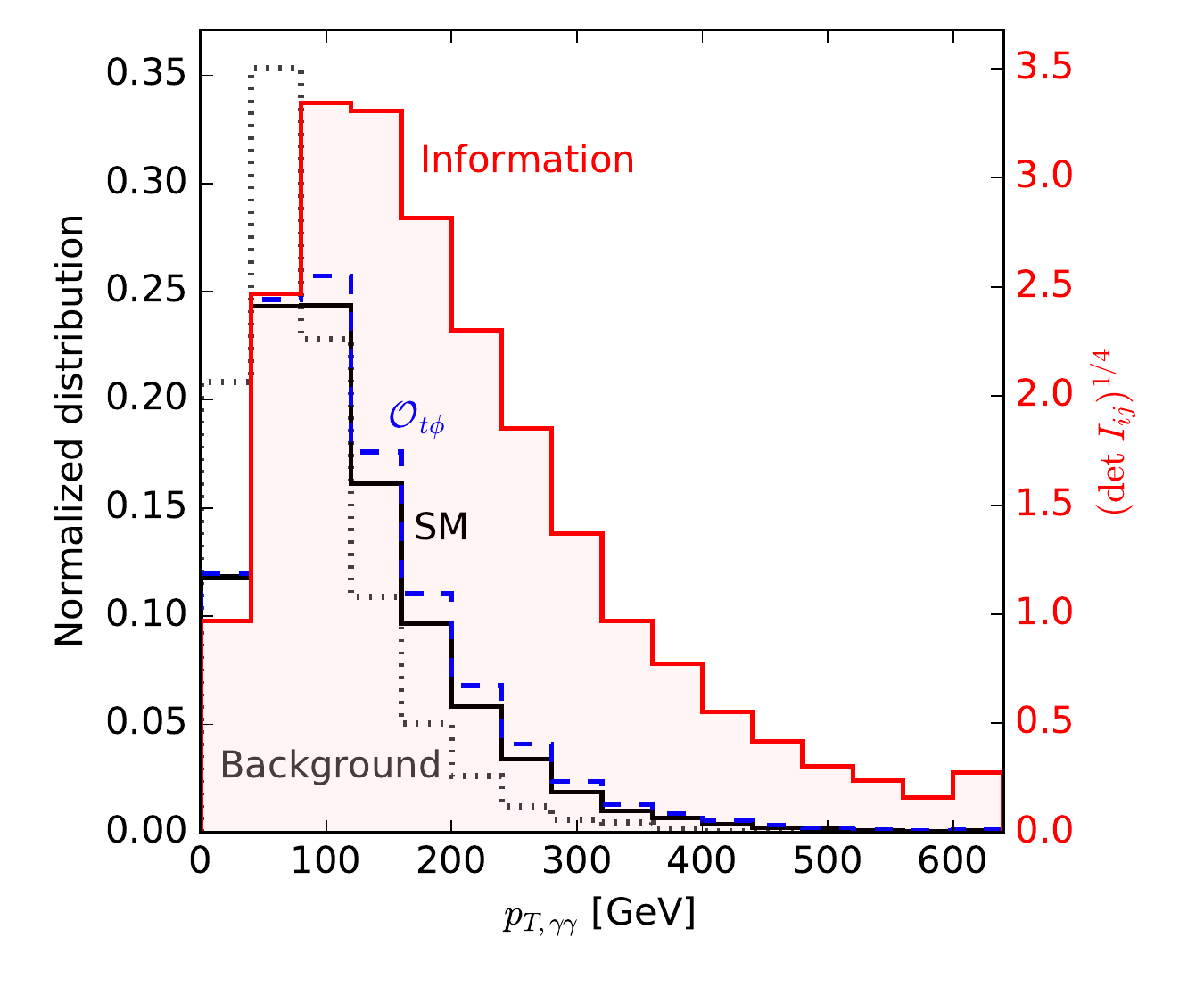}%
  \includegraphics[height=0.45 \textwidth]{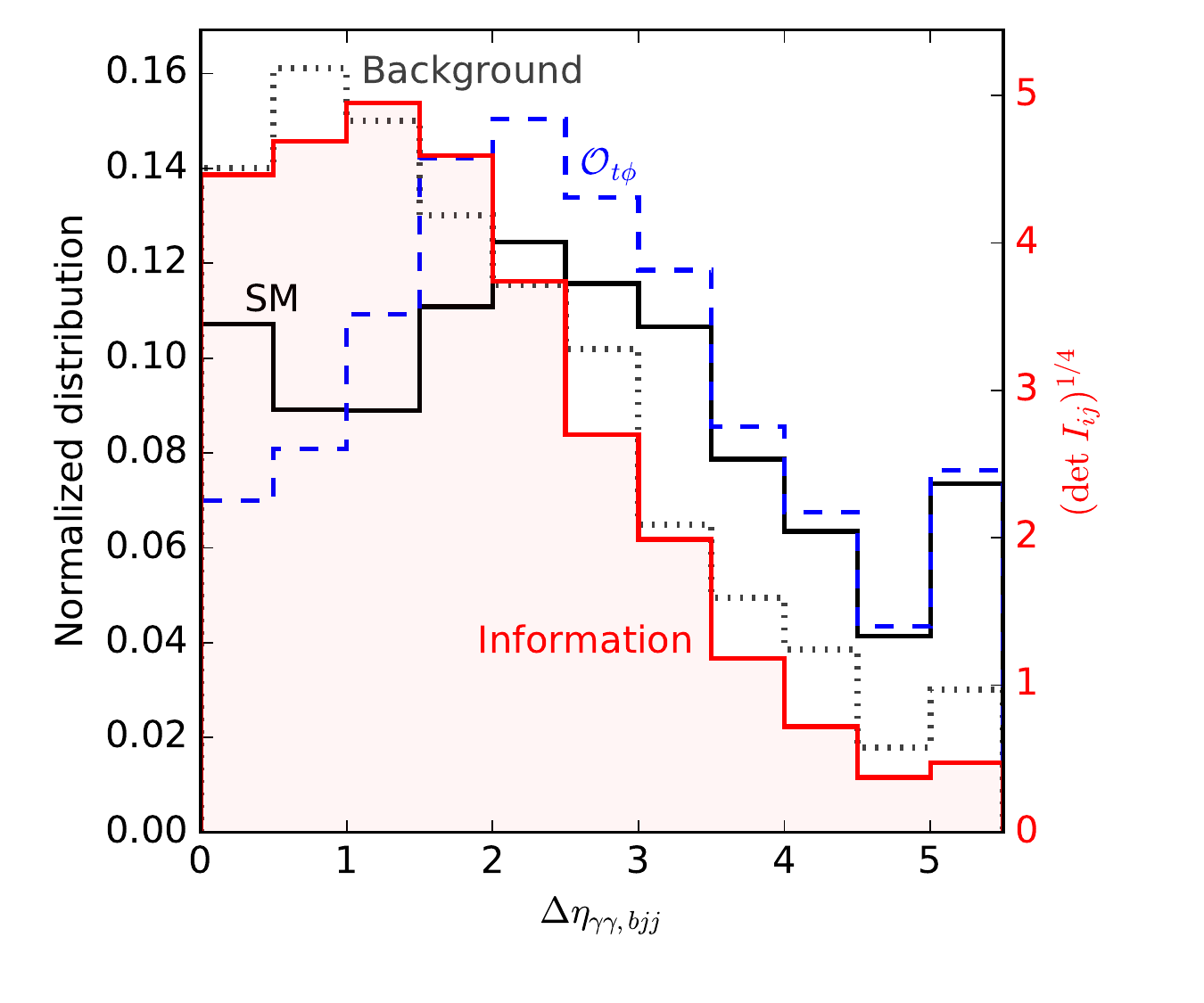}%
  \caption{Distribution of the Fisher information in the Higgs plus
    single top channel (shaded red). We also show the normalized SM
    signal (solid black) and single-top background (dotted grey)
    rates. The dashed blue line shows the effect of
    $f_{t \phi} \, v^2 / \Lambda^2 = 0.2$. The last bin is an overflow
    bin.}
  \label{fig:th_differential_information}
\end{figure}

In Fig.~\ref{fig:th_differential_information} we show the distribution
of this information over phase space. More distributions are shown in
Appendix~\ref{sec:additional_plots}. As expected, the information is
concentrated in the $m_{\gamma \gamma} \sim m_H$ peak and in the
high-energy tails of transverse momenta. Studying angular correlations
between the Higgs system and the top decay products, we find that the
region $\Delta \eta_{ \gamma \gamma, bjj} \lesssim 3$ contains a lot
of discrimination power.

\subsection{Information in distributions}

In a next step, we compare this full information to the reduced
information in one-dimensional and two-dimensional distributions of
kinematic observables. We now require harder cuts
\begin{align}
  p_{T,j_1} > 50~\gev \qquad 
  p_{T,\gamma} > 50,~30~\gev \qquad
  122~\gev < m_{\gamma \gamma} < 128~\gev \,,
  \label{eq:th_cuts}
\end{align}
which reduces the background to the level of the signal. We then
analyze the distributions~\cite{top_higgs,Kling:2012up}
\begin{itemize}[label=\raisebox{0.1ex}{\scriptsize$\bullet$}]
\item $p_{T,\gamma_1}$ with bin size 25~GeV up to 400~GeV and an
  overflow bin;
\item $m_{\gamma\gamma}$ with bin size 1~GeV in the allowed range of
  $123~...~127~\gev$;
\item $p_{T,\gamma \gamma}$ with bin size 40~GeV up to 600~GeV and an
  overflow bin;
\item $\Delta \phi_{\gamma \gamma}$ with bin size $\pi/10$;  
\item $p_{T,j_1}$ with bin size 40~GeV up to 400~GeV and an
  overflow bin;
\item $p_{T,b}$ with bin size 40~GeV up to 400~GeV and an
  overflow bin;
\item $p_{T,bjj}$ with bin size 40~GeV up to 600~GeV and an
  overflow bin;
\item $\Delta \phi_{\gamma \gamma, b}$ with bin size $\pi / 10$;
\item $\Delta \eta_{\gamma\gamma, b}$ with bin size $0.5$ up to $5.0$ and an
  overflow bin;
\item $m_{\gamma \gamma bjj}$ with bin size 100~GeV up to 1500~GeV and an
  overflow bin;
\item $p_{T,\gamma \gamma bjj}$ with bin size 40~GeV up to 400~GeV and an
  overflow bin;
\item $\Delta \phi_{\gamma \gamma, bjj}$ with bin size $\pi / 10$;
\item $\Delta \eta_{\gamma\gamma, bjj}$ with bin size $0.5$ up to $5.0$ and an
  overflow bin.
\end{itemize}\bigskip 

\begin{figure}
  \includegraphics[height=0.45 \textwidth]{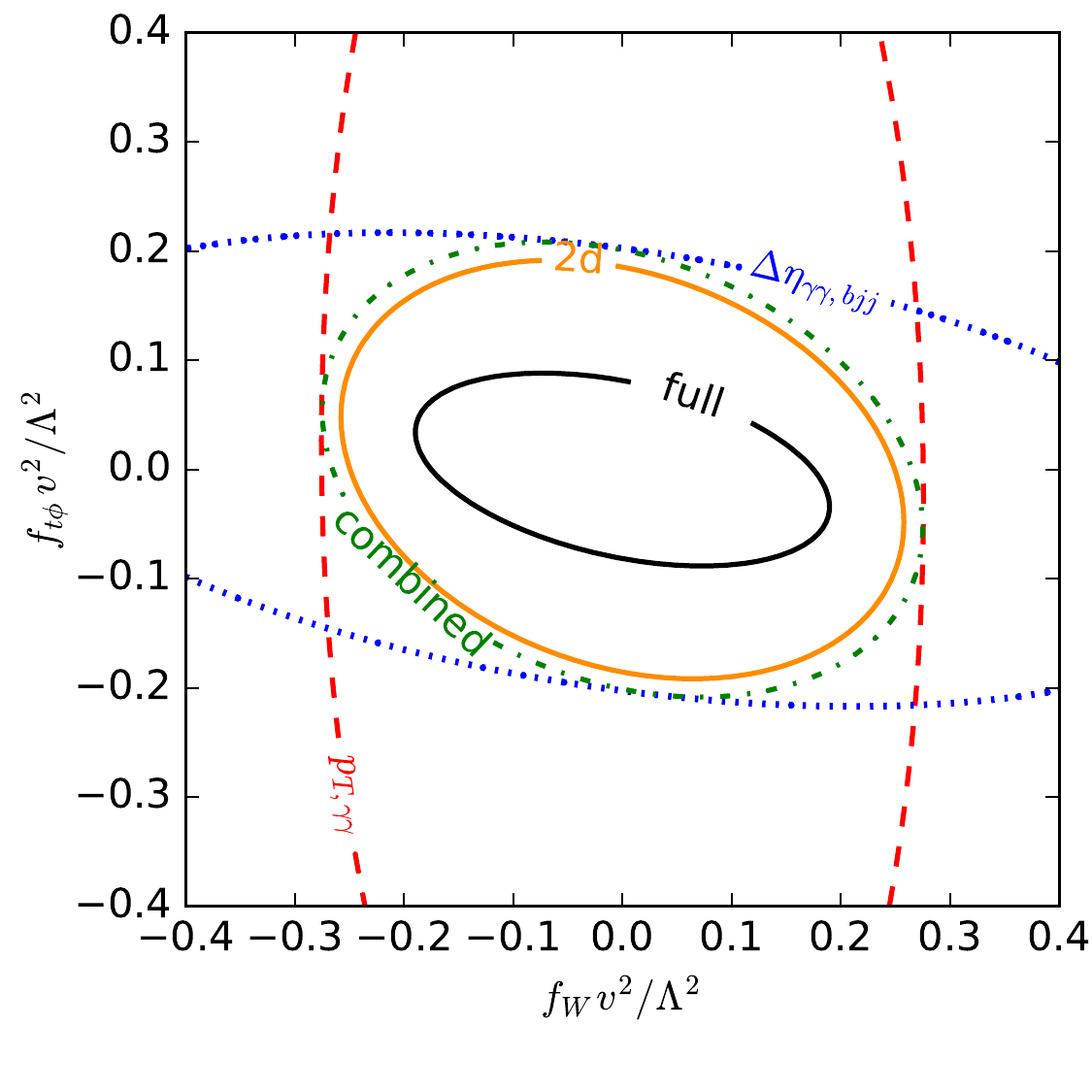}
  \caption{Information from histograms compared to the full
    information (black), shown as contours
    $d_{\text{local}}(\boldg ; \boldzero) = 1$. We include
    $p_{T,\gamma \gamma}$, $\Delta \eta_{\gamma\gamma, bjj}$, their naive combination assuming
    no mutual information, and their two-dimensional histogram. The
    $g_i$ not shown are set to zero.}
  \label{fig:th_histograms_contours}
\end{figure}

As in the WBF case, different observables probe different Wilson
operators. Figure~\ref{fig:th_histograms_contours} shows that
the di-photon transverse momentum constrains mostly the $\ope{W}$
direction, while the rapidity separation between the Higgs and top
systems is more sensitive to $\ope{t\phi}$.

\begin{figure}
  \includegraphics[height=0.6 \textwidth]{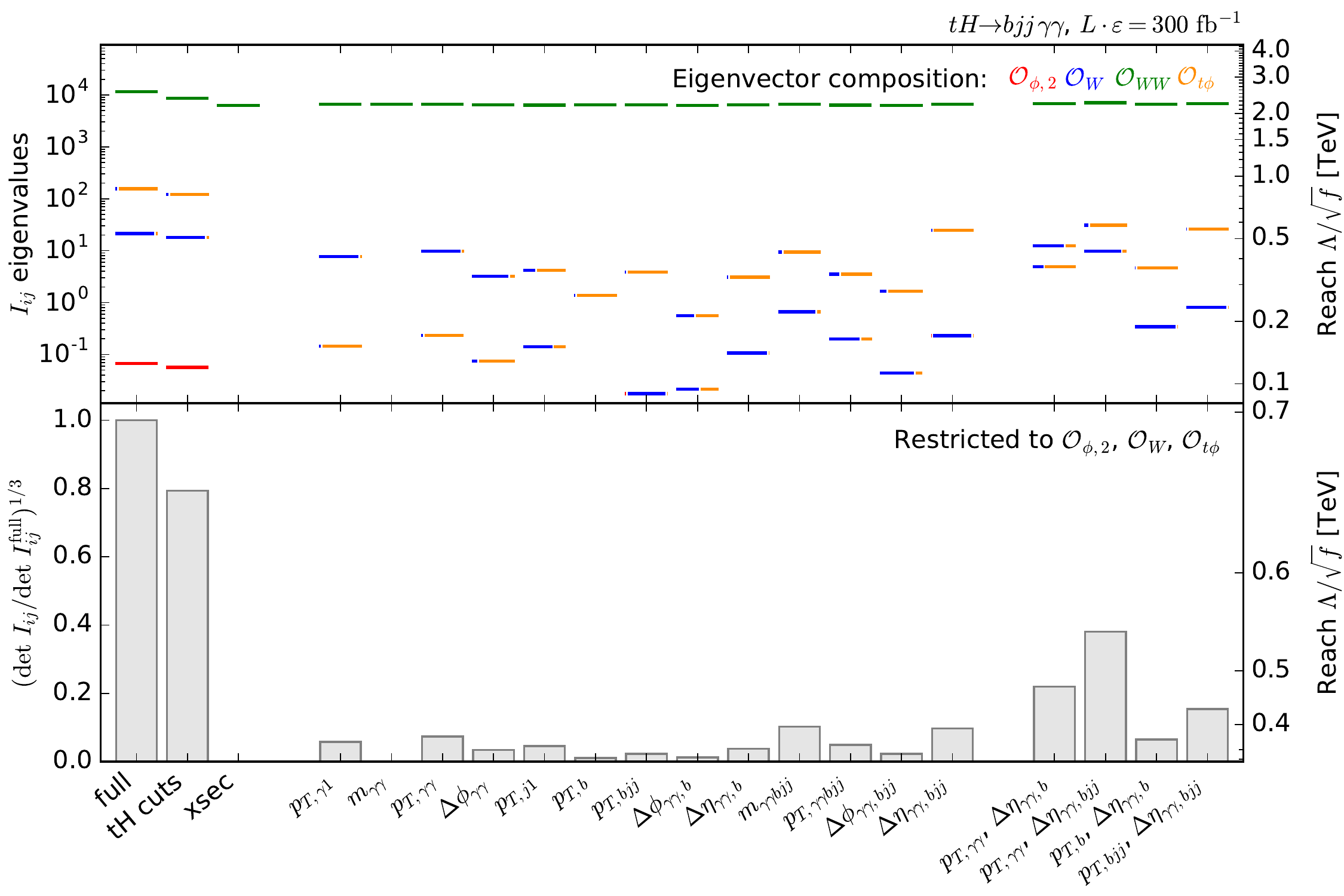}
  \caption{Fisher information for the Higgs plus single top channel
    exploiting the full phase space, after the cuts in
    Eq.\;\eqref{eq:th_cuts}, and for several kinematic distributions.
    The top panel shows the eigenvalues, the colors denote the
    composition of the corresponding eigenvectors. The right axis
    translates the eigenvalues into a new physics reach for the
    corresponding combination of Wilson coefficients.  In the bottom
    panel we show the determinants of the Fisher information
    restricted to $\ope{\phi,2}$, $\ope{W}$, and $\ope{t\phi}$,
    normalized to the full information. Again, the right axis
    translates them into a new physics reach.}
\label{fig:th_histograms_comparison}
\end{figure}

In Fig.~\ref{fig:th_histograms_comparison} we compare the eigenvalues,
eigenvectors and determinants of the information matrices in all of
the above distributions. We confirm that the photon observables mostly
probe changes in the Higgs-gauge coupling from $\ope{W}$, while a
rescaled top Yukawa will be visible in the properties of the top decay
products. Distributions of the properties of the $b$ jet consistently
contain significantly less information than the corresponding
distributions for the reconstructed top system. The rapidity
difference between the $\gamma \gamma$ system and the reconstructed
top provides a particularly good probe of this
operator~\cite{top_higgs}. Combining this variable with the transverse
momentum of the $\gamma \gamma$ system we can probe new physics scales
in the $\ope{\phi,2}$-$\ope{W}$-$\ope{t\phi}$ space around 550~GeV,
compared to 700~GeV for the fully differential cross section. This
corresponds to almost three times more data under our simplifying
assumptions.

\section{Conclusions}
\label{sec:conclusions}

We have used information geometry to calculate the maximum sensitivity
of Higgs measurements to dimension-6 operators, to understand the
structure of the observables, and to discuss how to improve these
measurements. Our approach is based on the Fisher information matrix,
which according to the Cram\'er-Rao bound defines the maximum
precision that can be achieved in a measurement. Unlike traditional
multivariate analysis techniques, it is designed for continuous,
high-dimensional parameter spaces like effective field theories. We
have demonstrated how the Fisher information can be reliably
calculated using Monte-Carlo techniques.

Going beyond global statements, the Fisher information can be studied
differentially to understand how the discriminating power is
distributed over phase space, which helps guide event selection
strategies.  Moreover, we can also calculate the information contained
in subsets of kinematic distributions. This helps us determine
which observables are the most powerful, and allows us to compare the
constraining power in conventional analyses with one or two variables
to that in more complex multivariate analyses.\bigskip

Our first testing ground was Higgs production in weak boson fusion
with decays into a tau pair or into four leptons. Crucial information
comes from the high-energy tails as well as from angular correlations
between jets. Decay kinematics hardly adds any information, since the
momentum flow is limited by the Higgs mass and gauge invariance forces
us to include operators with a momentum dependence.  Tight cuts on the
rapidity separation of the tagging jets throw away a large amount of
discrimination power. Under idealized conditions, conventional
analyses based on a simple event selection and standard kinematic
distributions can probe new physics scales around $900~\gev$ in the
early phase of Run~II. Multivariate analyses have the potential to
significantly enhance the sensitivity and probe new physics scales of
up to $1.2~\tev$.

In Higgs production with a single top we find that kinematic
properties of the Higgs decay products and observables in the top
system provide orthogonal information. The transverse momenta of the
di-photon system as well as the rapidity separation of the
$\gamma \gamma$ and $bjj$ systems are powerful observables. But even
with HL-LHC data these distributions are only sensitive to new physics
scales around $550~\gev$, while a multivariate analysis might be able
to probe scales up to $700~\gev$.\bigskip

To summarize, information geometry provides a powerful and intuitive
tool that can help understand the phenomenology of models with a
continuous, high-dimensional parameter space, and in turn can be used
to optimize measurement strategies.  We have demonstrated this
approach in different Higgs channels for dimension-6 operators, but
it can easily be translated to other processes and models.

\subsubsection*{Acknowledgments}

We would like to thank Juan Gonzalez-Fraile for many fruitful
discussions and sharing his model file for some of the dimension-6
operators. We are grateful to Peter Schichtel for helping us to set up
and use \toolfont{MadMax}.  We would also like to acknowledge Ben
Allanach's early interest in pursuing information geometry in the
context of supersymmetry.

JB is funded by the DFG through the Graduiertenkolleg \emph{Particle
  physics beyond the Standard Model} (GRK~1940). KC is supported
through NSF ACI-1450310 and PHY-1505463. The work of FK is supported
by NSF under Grant PHY-1620638. TP is supported by the DFG
Forschergruppe \emph{New Physics at the LHC} (FOR~2239). The authors
acknowledge support by the state of Baden-W\"urttemberg through bwHPC.

\clearpage

\appendix
\section{Additional results}
\label{sec:appendix}

\subsection{A simple example}
\label{sec:simple_example}

As a simple example we study the Fisher information in a number of
rates $n_c$ measured in various Higgs channels $c$. In the absence of
systematic uncertainties they follow Poisson statistics,
\begin{align}
  f (\mathbf{n} |\boldsymbol{\nu}) 
  = \prod_c \Pois (n_c | \nu_c) 
  = \prod_c \, \frac{\nu_c^{n_c} e^{-\nu_c}}{n_c!} \, .
\end{align}
We can calculate the Fisher information in terms of the Poisson mean
$\boldsymbol{\nu}$ as
\begin{align}
  \pder { \log f} {\nu_c} 
  = \frac {n_c} {\nu_c} - 1 \qqquad 
  \frac { \partial^2 \log f} {\partial \nu_c \partial \nu_{c'}} = - \frac {\delta_{c c'} \, n_c} {\nu_c^2} \qqquad
  I_{c c'} \equiv - E \left[ \frac { \partial^2 \log f} {\partial \nu_c \partial \nu_{c'}} \middle| \boldsymbol{\nu} \right] = \frac {\delta_{c c'} } {\nu_c} \,.
\end{align}
If we express the expected count rates in terms of model parameters
$g_i$, the Fisher information becomes
\begin{align}
  I_{ij}  = \sum_c \frac 1 {\nu_c} \, \pder {\nu_c} {g_i} \, \pder {\nu_{c}} {g_j} \,.
  \label{eq:simple_example_information}
\end{align}

The matrix $\partial \nu_c / \partial g_i$ is determined by selection
requirements, detector acceptance, and efficiencies.  In the $\kappa$
framework that only scales cross sections and branching ratios, the
matrix $\partial \nu_c / \partial g_i$ is trivial to calculate in
closed form.  For each channel this matrix is singular, which means it
measures one direction in parameter space and is blind to all
others. At least as many channels as parameters are required to make
the combined information in Eq.\;\eqref{eq:simple_example_information}
non-singular and remove all blind directions (assuming the channels do
not provide degenerate information, \ie the same eigenvectors in the
Fisher information).\bigskip

For illustration, we consider the case where we want to
measure one coupling $g$ in one channel with the expected number of
events
\begin{align}
\nu = L \left( \sigma_S + \sigma_B \right) 
    = L g^2 \sigma_0 + L \sigma_B \; .
\end{align}
The Fisher information is then
\begin{align}
I = 4 L \, \frac {g^2 \sigma_0^2 } {g^2 \sigma_0 + \sigma_B} 
  = \frac{4 L}{g^2} \, \frac {\sigma_S^2 } {\sigma_S + \sigma_B} \, .
\end{align}
According to the Cram\'er-Rao bound, the standard deviation of any
unbiased estimator $\hat{g}$ is at least
\begin{align}
\frac{\Delta \hat{g}}{g} \geq \frac 1 {g \, \sqrt{I}} = 
\frac{1}{2 \, \sqrt{L}} \, \frac {\sqrt{\sigma_S +\sigma_B}} {\sigma_S} \,.
\end{align}
The three terms show how the sensitivity to $g$ profits from the
square in the cross section, the square-root dependence on the
statistics, and the dependence on the signal-to-background ratio.

\subsection{The MadFisher algorithm}
\label{sec:algorithm}

We calculate the Fisher information in Eq.\;\eqref{eq:fisher_rates}
with Monte-Carlo methods. With
\begin{equation}
  \int \! d x \; f^{(1)} (x) \to \sum_{\text{events k}} \Delta \sigma_k / \sigma
\end{equation}
we find
\begin{equation}
  I_{ij} (\boldg)
  = \sum_{\text{events } k} \frac{L}{\Delta \sigma_k (\boldg)}
  \pder { \Delta \sigma_k (\boldg)} {g_i} \,
  \pder { \Delta \sigma_k (\boldg)} {g_j} \,,
  \label{eq:information_from_events}
\end{equation}
requiring the differential cross sections and their derivatives as
input.

We first generate event samples for a number of benchmark parameters
with \toolfont{MadMax}~\cite{madmax2}. This add-on to
\toolfont{MadGraph~5}~\cite{madgraph} allows us to simultaneously
calculate differential rates for different parameter points using the
same phase-space grid. Our \toolfont{FeynRules}~\cite{feynrules} model
file of the relevant dimension-6 operators does not truncate operator
effects at $\ord{1/\Lambda^2}$. \toolfont{MadMax} requires fixed
renormalization and factorization scales, which we set following
Ref.~\cite{yr4}. To keep the calculation times manageable, we restrict
some processes to the dominant sub-processes, for instance to
initial-state $u$ and $d$ quarks in the WBF case.  We then normalize
the Higgs rates to the LHC HXS WG recommendations for the total cross
section~\cite{yr4}, calculating the effect of the different acceptance
regions with \toolfont{MadGraph~5}. Background processes are simply
rescaled to \toolfont{MadGraph} predictions.

A morphing technique allows us to calculate the differential cross
sections and their derivatives at arbitrary positions in parameter
space~\cite{morphing}. The effect of $\ope{\phi,2}$ and systematic
rate uncertainties (see Sec.~\ref{sec:systematics}) is taken into
account analytically. The contributions from the other operators are
decomposed into a number of basis components and can be exactly
reconstructed from a set of simulated benchmark points. Our example
processes require up to 70 such basis components (in the WBF
$H \to 4 \ell$ case).

We can then easily calculate the Fisher information according to
Eq.\;\eqref{eq:information_from_events}. Finally, global distances as in
Eq.\;\eqref{eq:distances} are calculated in analogy to free fall in
general relativity: a starting point and a set of directions in
parameter space define the initial conditions, from which we
numerically calculate distances along curves defined by the geodesic
equation.

\subsection{Additional distributions}
\label{sec:additional_plots}

\begin{figure}
  \includegraphics[height=0.45 \textwidth]{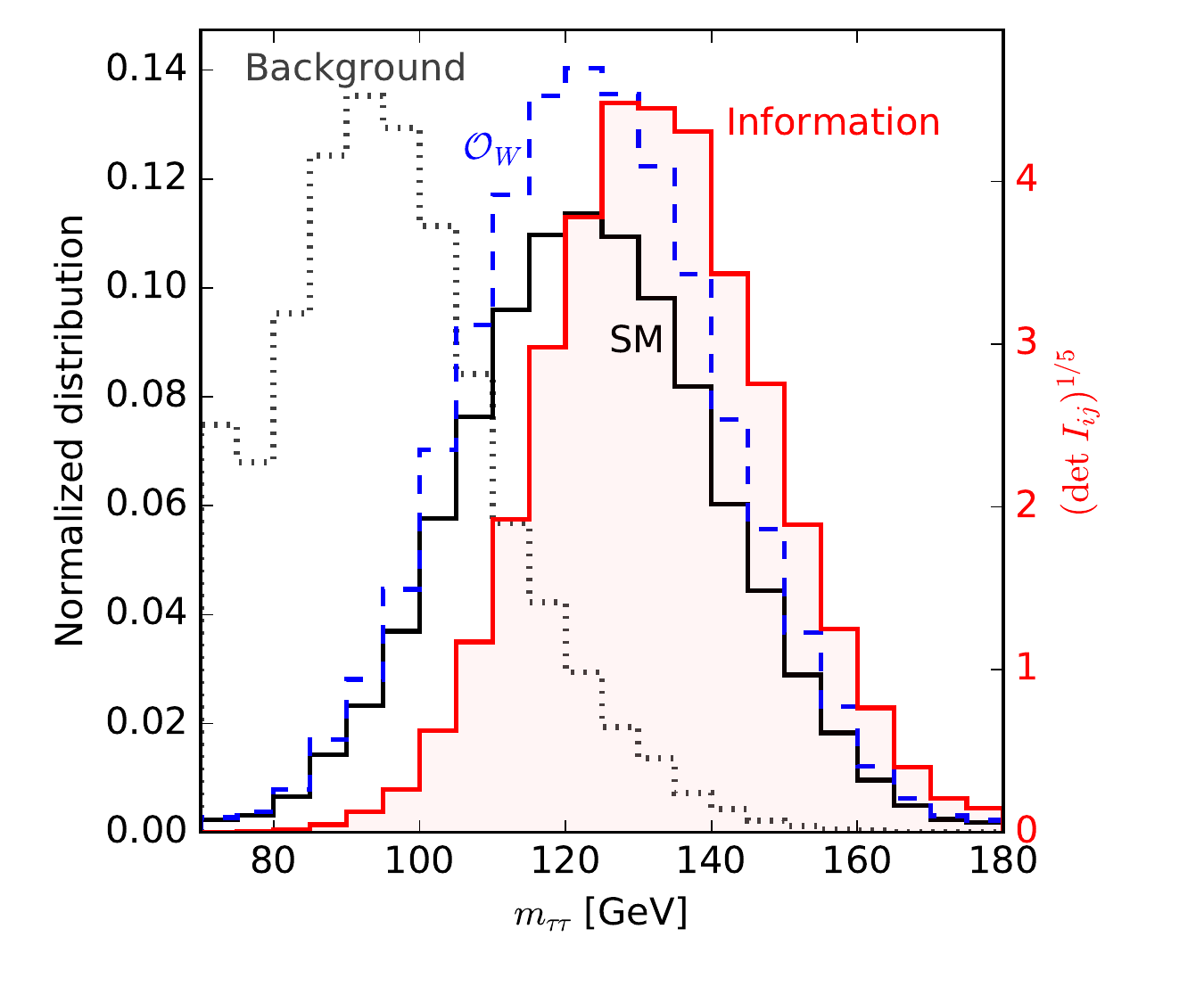}%
  \includegraphics[height=0.45 \textwidth]{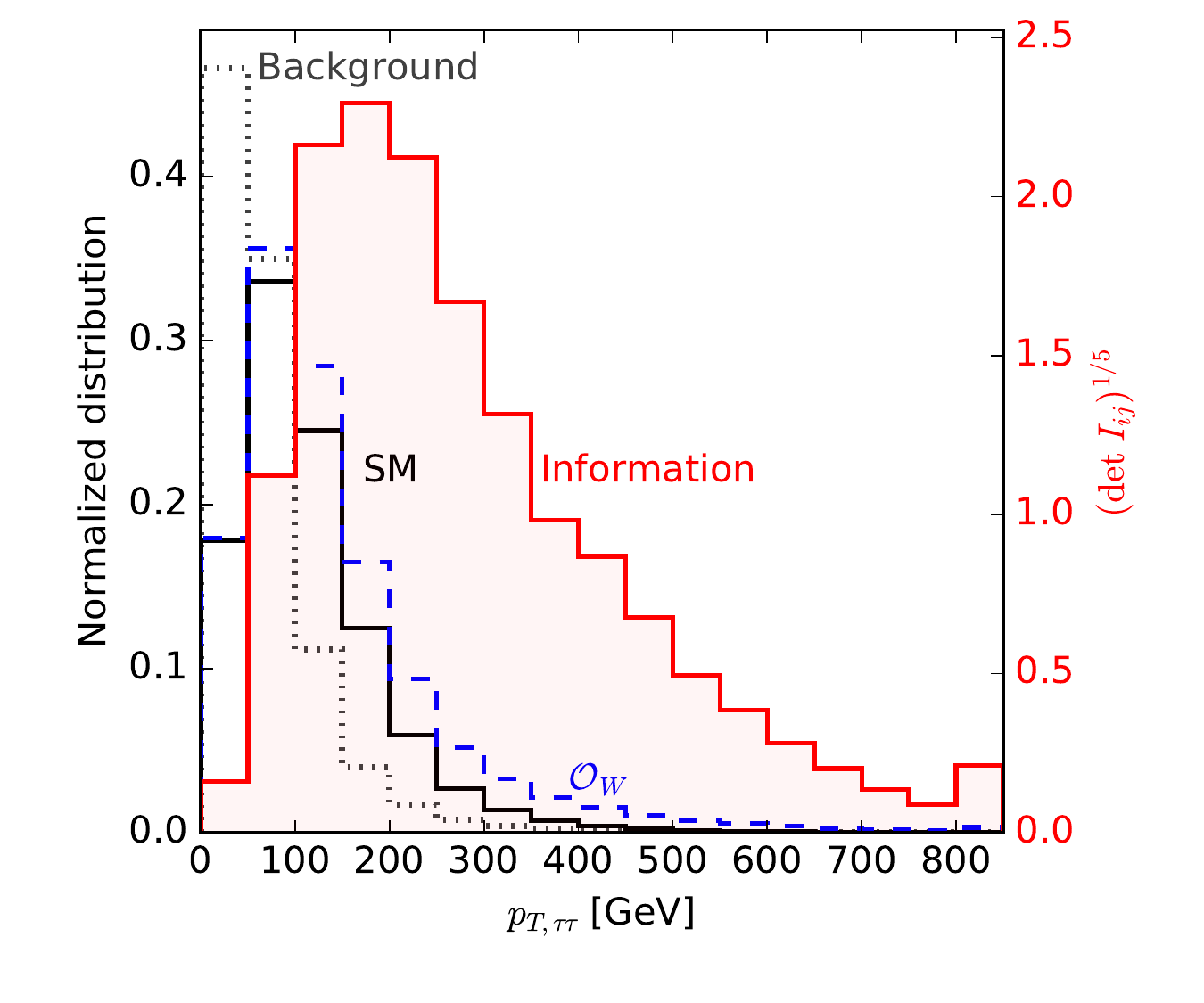}\\%
  \includegraphics[height=0.45 \textwidth]{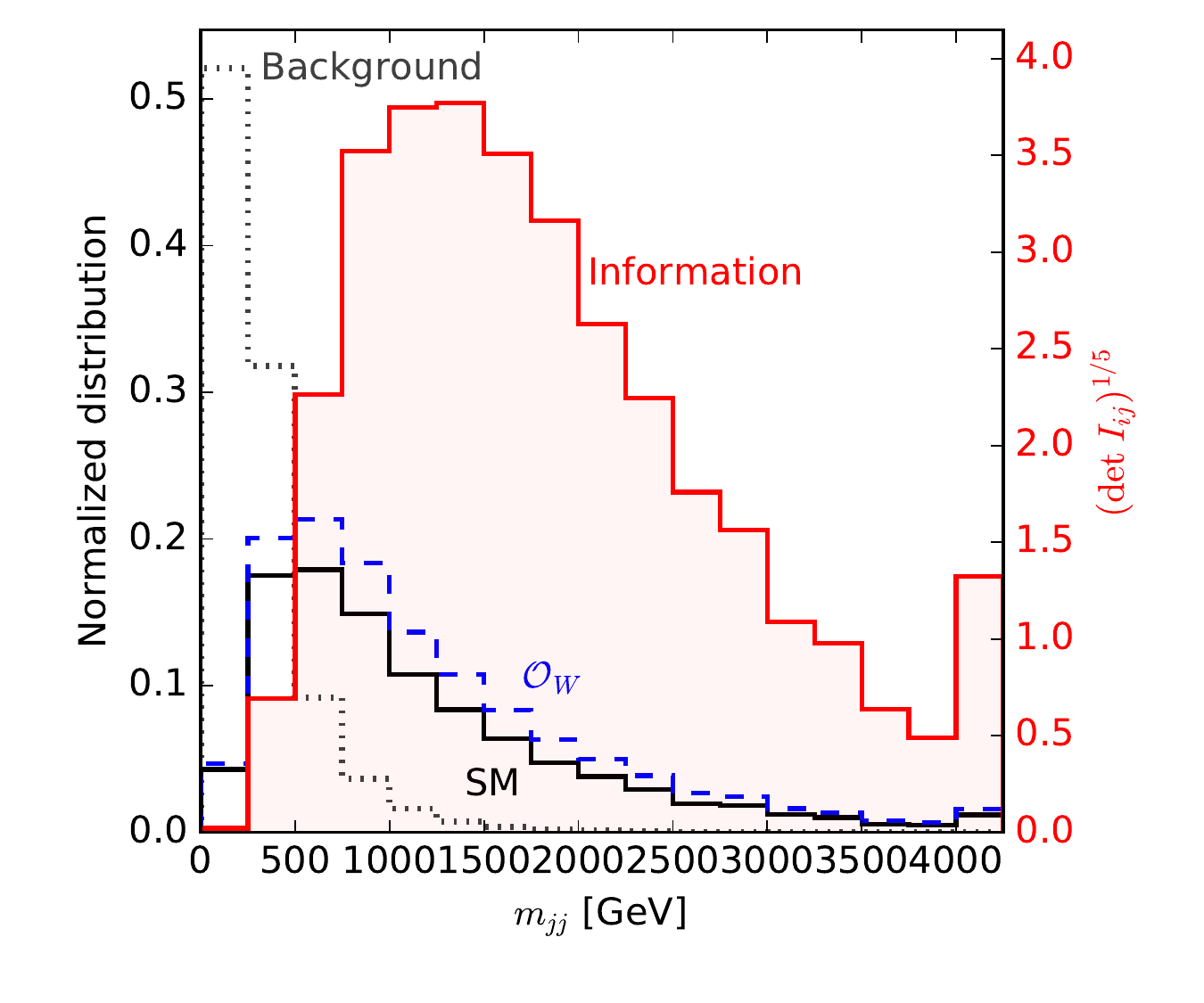}%
  \includegraphics[height=0.45 \textwidth]{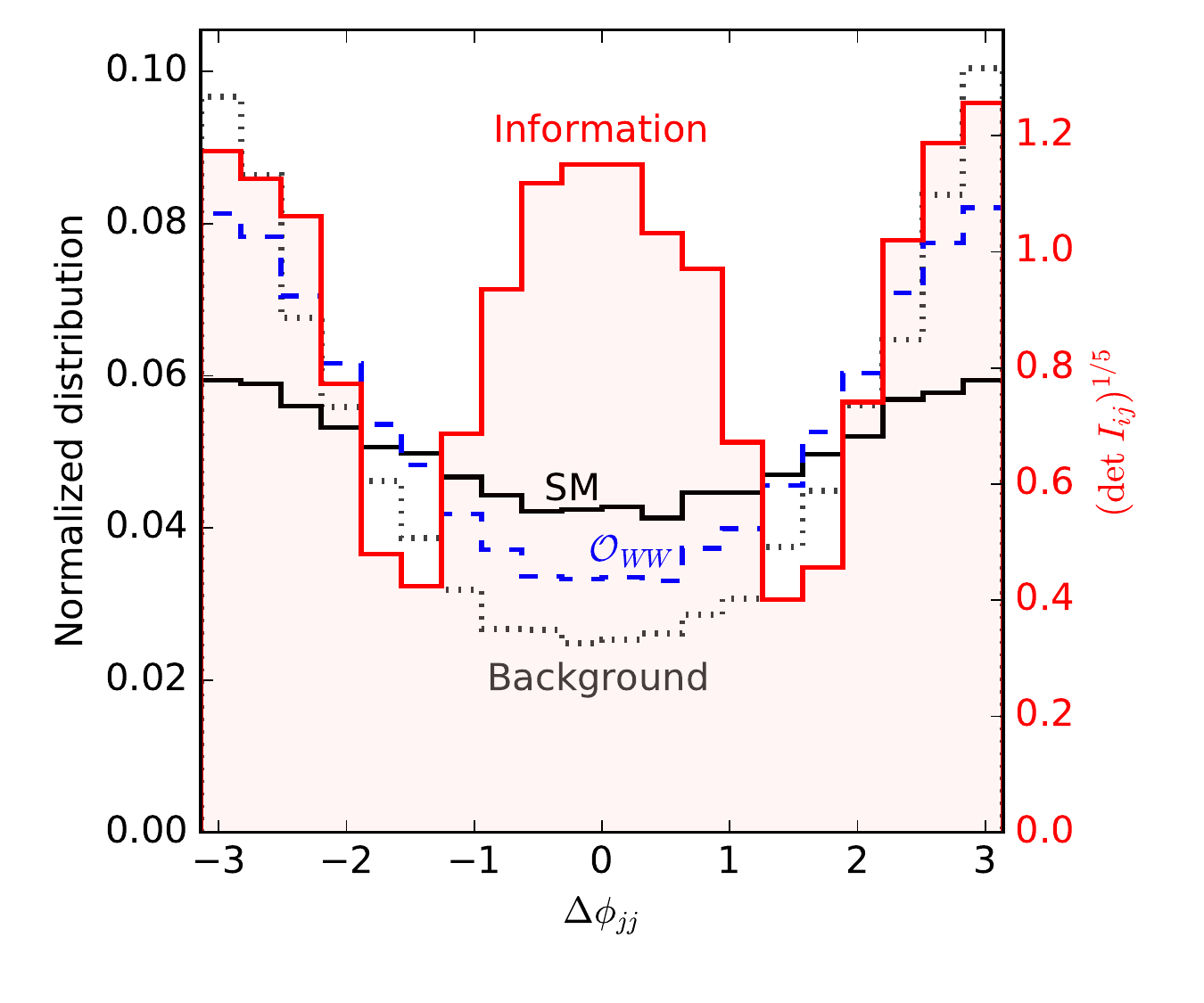}%
  \caption{Distribution of the Fisher information in the WBF
    $H \to \tau \tau$ channel (shaded red). We also show the
    normalized SM signal (solid black) and QCD $Z$+jets (dotted grey)
    rates. The dashed blue line shows the effect of an exaggerated
    $f_{W} \, v^2 / \Lambda^2 = 0.5$
    ($f_{WW} \, v^2 / \Lambda^2 = 0.5$ in the bottom right panel). The
    first (last) bins are underflow (overflow) bins.}
  \label{fig:wbf_tautau_differential_information_more}
\end{figure}

In Fig.~\ref{fig:wbf_tautau_differential_information_more} we show the
distribution of the differential information in the WBF
$H \to \tau \tau$ channel over various kinematic
variables. Figure~\ref{fig:th_differential_information_more}
contains similar distributions for the Higgs plus single top channel.

\begin{figure}
  \includegraphics[height=0.45 \textwidth]{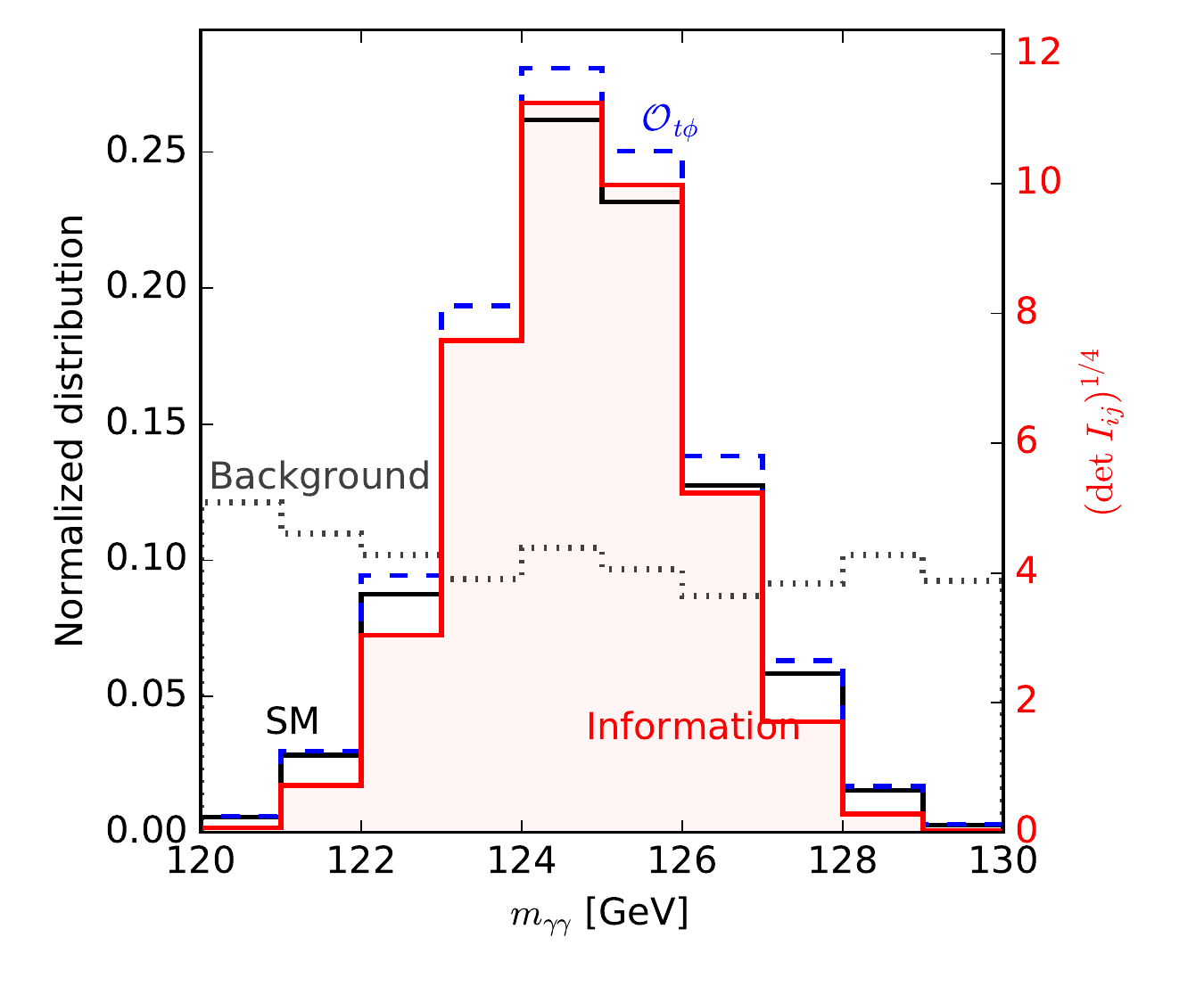}%
  \includegraphics[height=0.45 \textwidth]{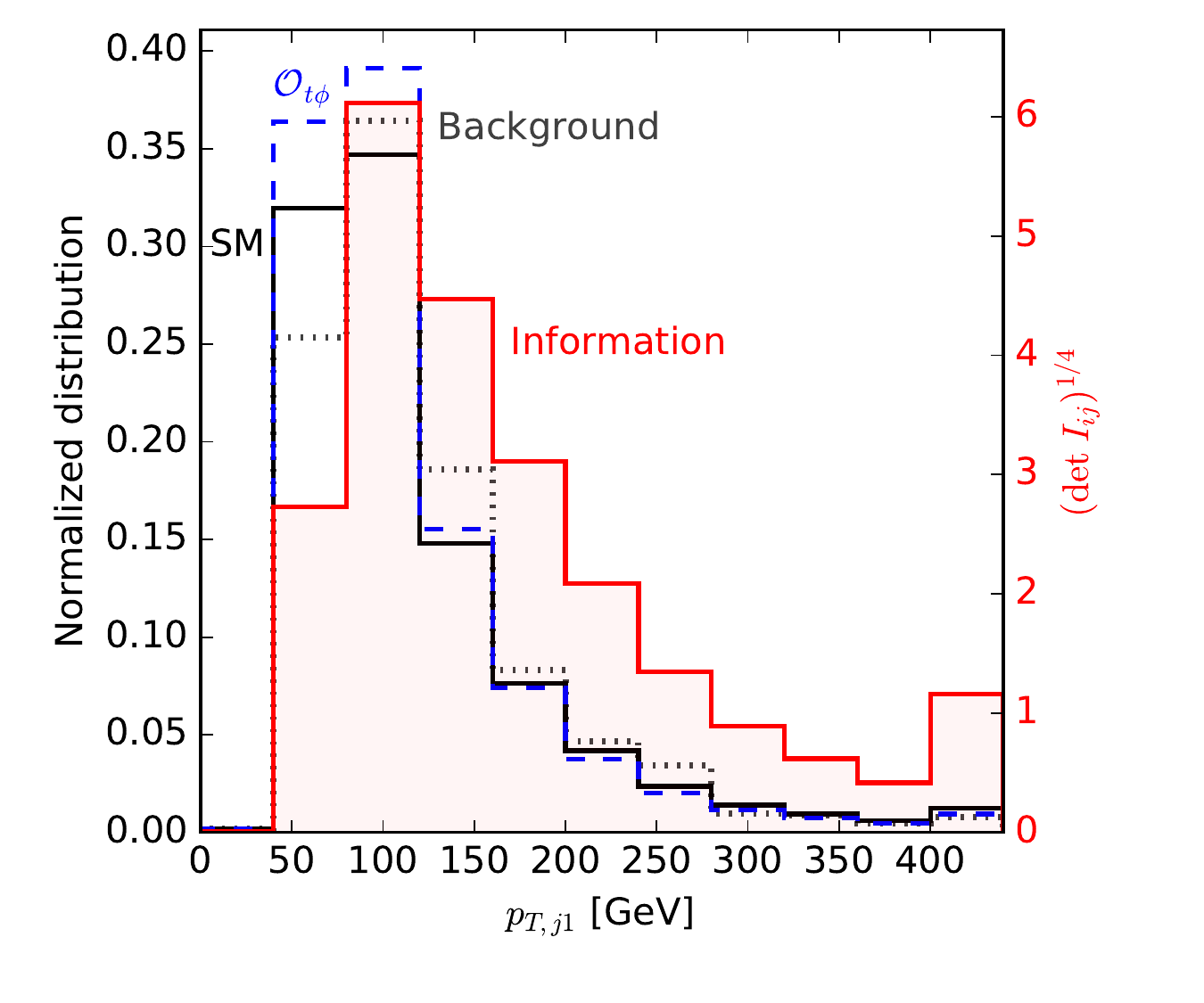}\\%
  \includegraphics[height=0.45 \textwidth]{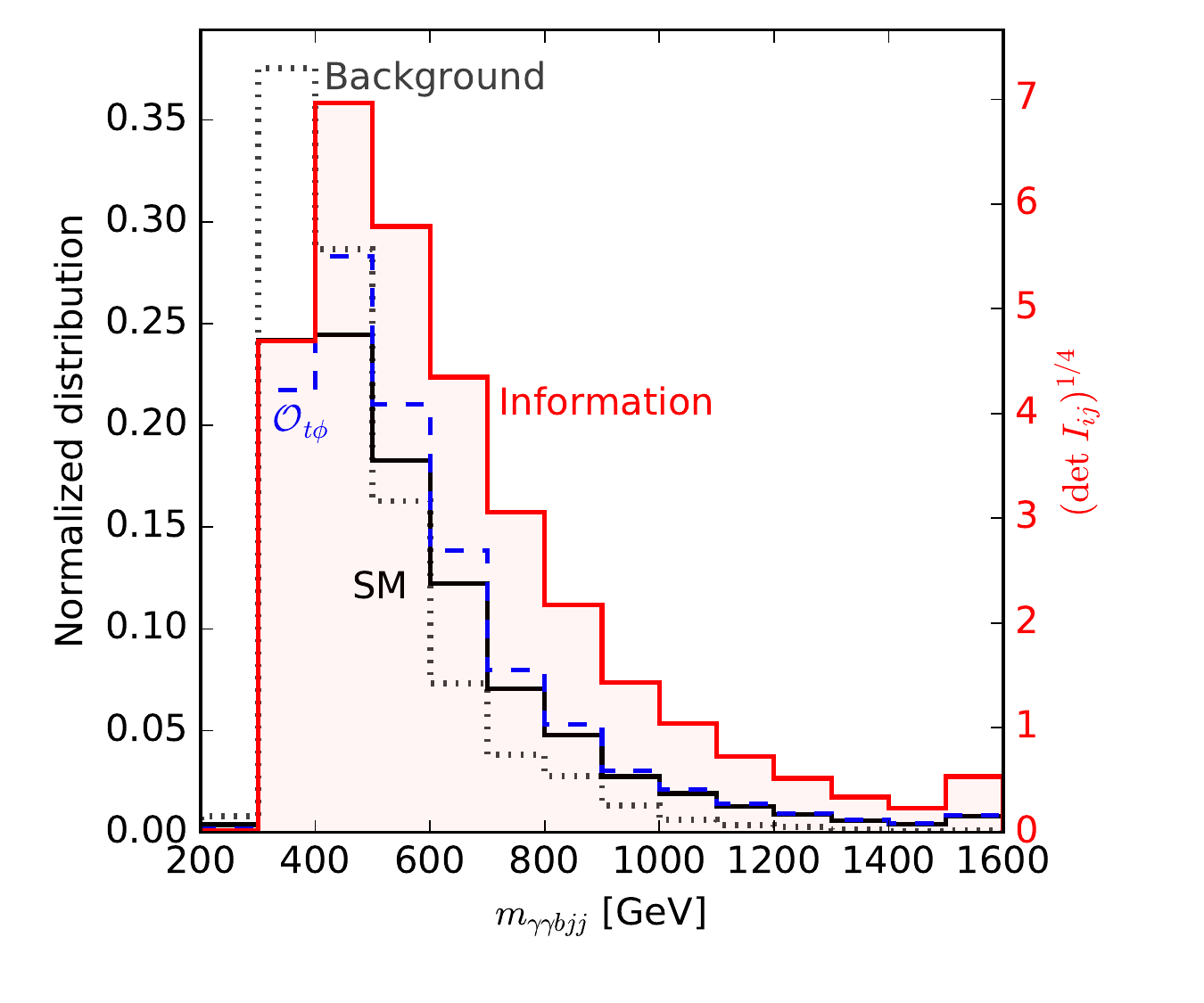}%
  \includegraphics[height=0.45 \textwidth]{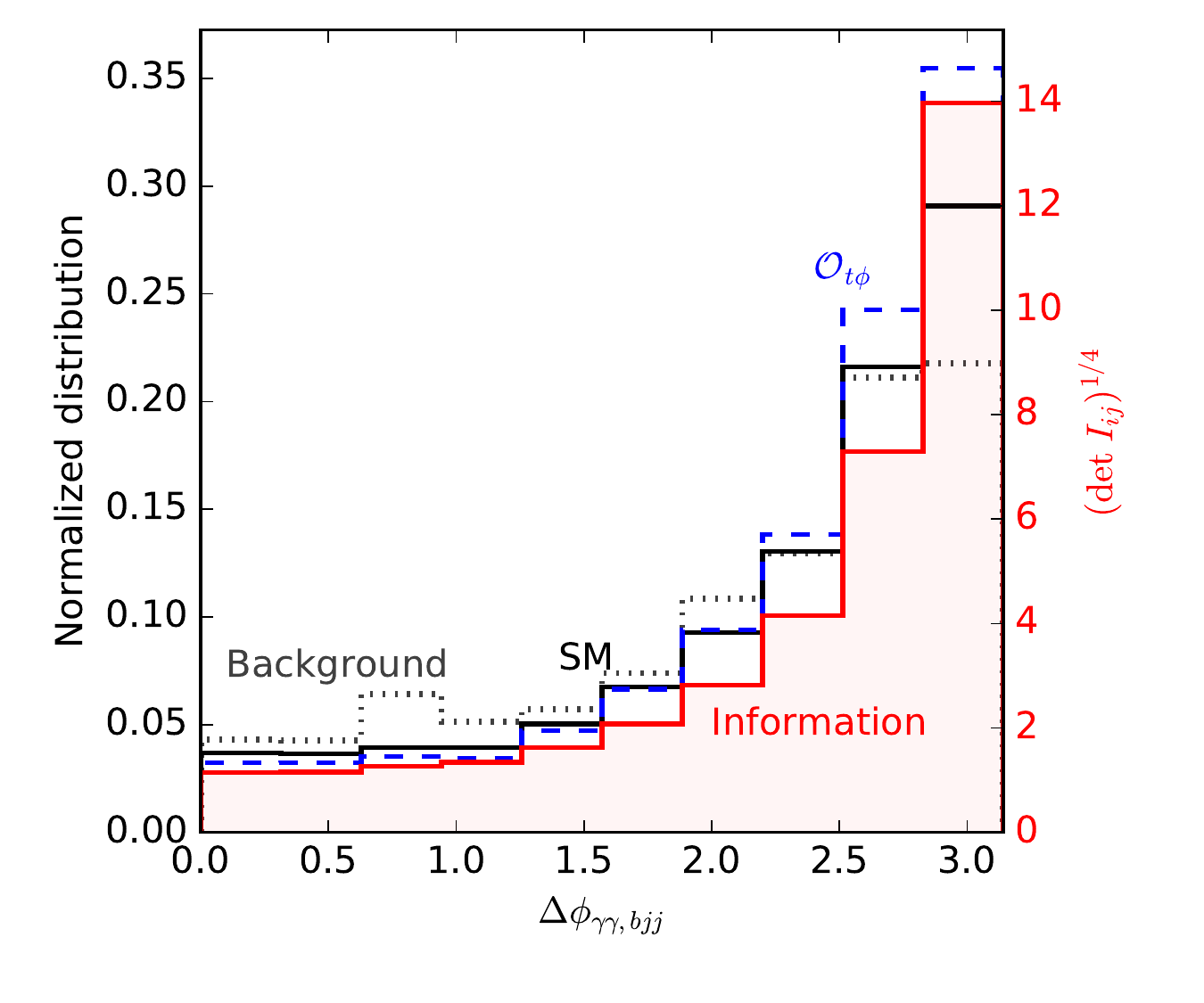}%
  \caption{Distribution of the Fisher information in the Higgs plus
    single top channel (shaded red). We also show the normalized SM
    signal (solid black) and single-top background (dotted grey)
    rates. The dashed blue line shows the effect of
    $f_{t \phi} \, v^2 / \Lambda^2 = 0.2$. The first (last) bins are
    underflow (overflow) bins.}
  \label{fig:th_differential_information_more}
\end{figure}

\subsection{Systematic uncertainties}
\label{sec:systematics}

Our information geometry approach can easily be extended to include
systematic and theory uncertainties. The parameter space then consists
of nuisance parameters $\nu_i$ in addition to the Wilson coefficients,
and constraint terms are added to the likelihood. If for instance the
$k$th parameter is a nuisance parameter with a Gaussian constraint
term with width $\sigma_k$, the additional term in the Fisher
information reads
\begin{align}
  I_{ij} (\boldg,\boldsymbol{\nu}) = \dots + \frac {\delta_{ik} \delta_{jk}} {\sigma_k^2} \,.
\end{align} 
This also applies to a log-normal constraint through reparametrization
of $\nu$.  Local and global distances now refer to combined theory and
nuisance parameters $(\boldg,\boldsymbol{\nu})$.

We define a profiled local distance between two points $\boldg_b$ and
$\boldg_a$
\begin{align}
  d_{\text{profiled}} (\boldg_b, \boldg_a)
  = \min_{\boldsymbol{\nu}}   \; d_{\text{local}}( (\boldg_b,\boldsymbol{\nu}) ;
  (\boldg_a , \mathbf{0} ) ) \,.
\end{align}
Equivalently, we can define a profiled Fisher information
matrix. Assuming the last parameter to be the only nuisance parameter,
the Fisher information matrix has the form
\begin{align}
  I_{ij} = \twomatc {I^{\text{theory}}} {\mathbf{m}} {\mathbf{m}^T} {n}
\end{align}
where $I^{\text{theory}}$ is the information matrix restricted to the theory
parameters, the vector $\mathbf{m}$ describes the mixing between
theory and nuisance parameter, and $n$ is the component that only
affects the nuisance parameters. Technically described by the parallel
projection of an ellipsoid, the projected Fisher information is given
by
\begin{align}
  I_{\text{profiled} \, ij} = I^{\text{theory}}_{ij} - \frac {m_i m_j} {n} \,.
\end{align}
\bigskip

\begin{figure}
  \includegraphics[height=0.45 \textwidth]{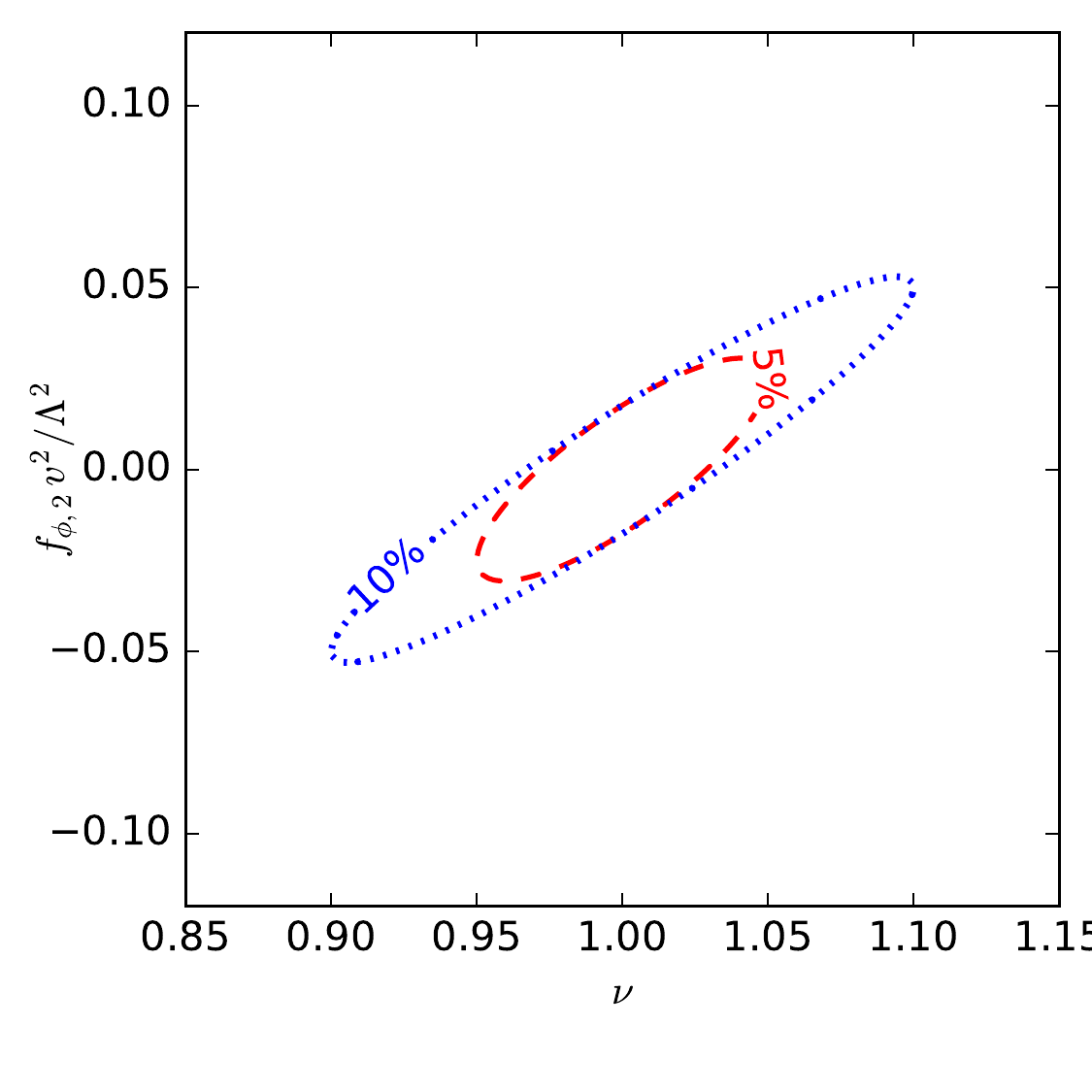}
  \hspace*{0.05\textwidth}
  \includegraphics[height=0.45 \textwidth]{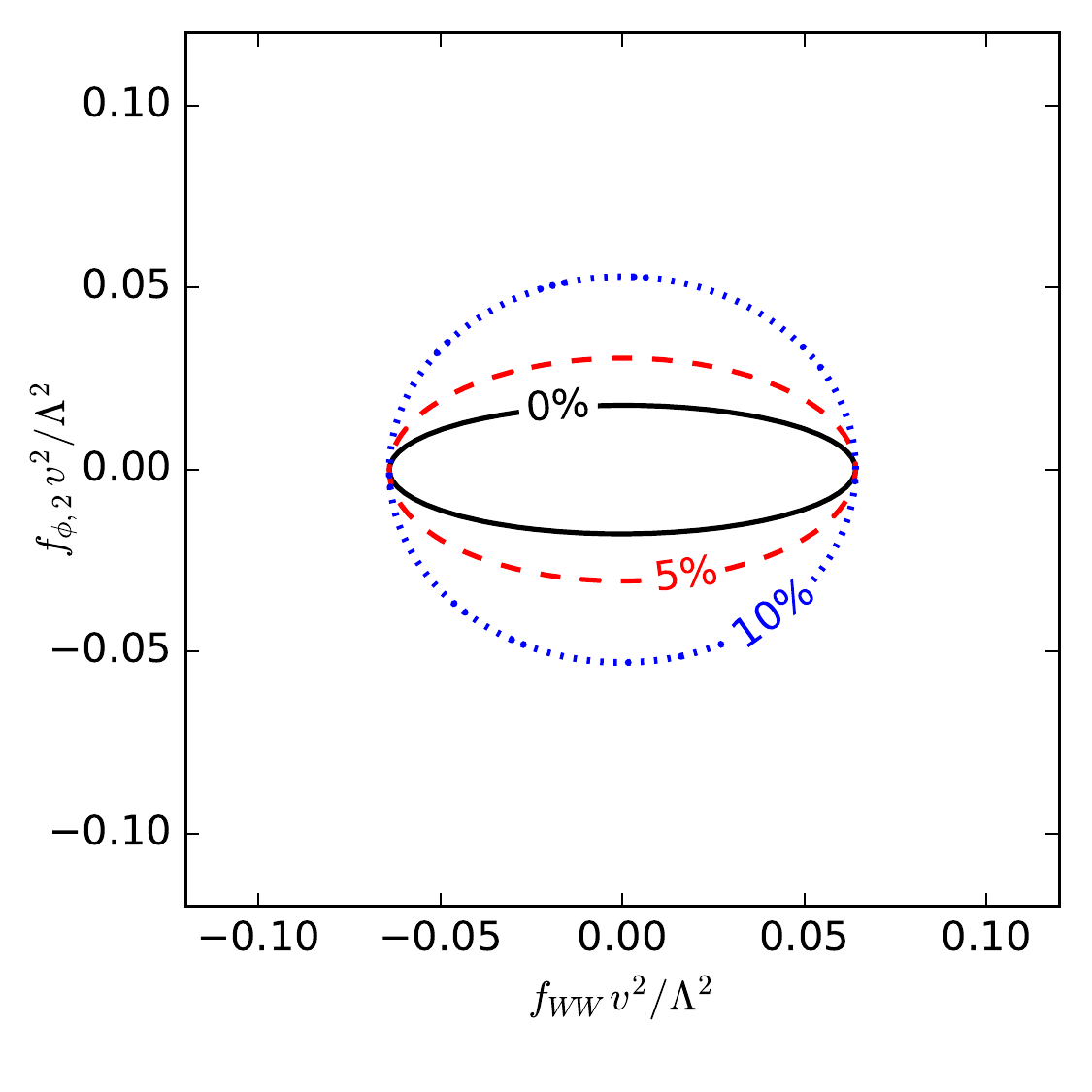}
  \caption{Effects of Gaussian uncertainties of $5\%$ and $10\%$ on
    the total signal rate. In the left panel we show the expected
    error ellipse $d_{\text{local}}( (g,\nu) ; \boldzero) = 1$ in
    the plane spanned by a physical parameter and the nuisance
    parameter $\nu$ rescaling the signal rate. In the right panel we
    show the error ellipses in the $\ope{W}$-$\ope{\phi,2}$ plane
    after profiling over this systematic uncertainty.}
  \label{fig:wbf_tautau_systematics}
\end{figure}

\begin{figure}
  \includegraphics[height=0.6 \textwidth]{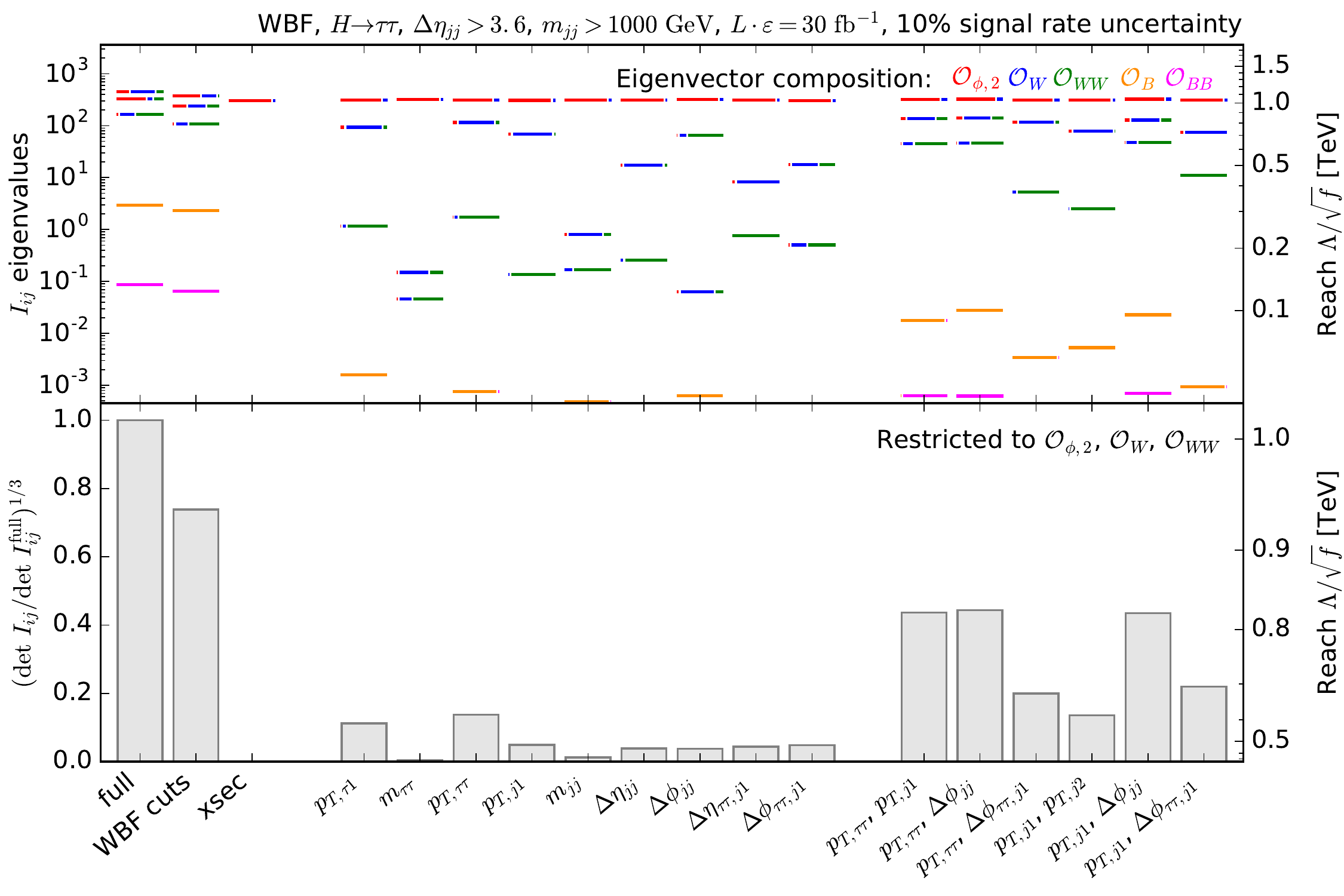}
  \caption{Fisher information for the WBF $H \to \tau \tau$ channel
    profiled over a $10\%$ signal rate uncertainty. We compare the
    information for the full phase space, after the cuts in
    Eq.\;\eqref{eq:wbf_tautau_wbfcuts}, and for several kinematic
    distributions.  The top panel shows the eigenvalues, the colors
    denote the composition of the corresponding eigenvectors. The
    right axis translates the eigenvalues into a new physics reach for
    the corresponding combination of Wilson coefficients.  In the
    bottom panel we show the determinants of the Fisher information
    restricted to $\ope{\phi,2}$, $\ope{W}$, and $\ope{WW}$,
    normalized to the full information. Again, the right axis
    translates them into a new physics reach.}
  \label{fig:wbf_tautau_systematics_comparison}
\end{figure}

We demonstrate this for WBF Higgs production in the $\tau \tau$ mode
in Fig.~\ref{fig:wbf_tautau_systematics}. We assign a $5\%$ or $10\%$
Gaussian uncertainty on the overall signal rate, representing for
instance missing higher orders, pdf or efficiency uncertainties. This
significantly reduces the information in the total rate, and thus
mostly the expected precision in the $\ope{\phi,2}$ direction.  In
Fig.~\ref{fig:wbf_tautau_systematics_comparison} we show how the
information in various distributions is affected by such an
uncertainty. The new physics reach in the $\ope{\phi,2}$ direction is
reduced by 800~GeV.

\subsection{Likelihood ratios and Fisher distance}
\label{sec:likelihood_ratio}

There is some subtlety in the relationship between 
standard likelihood ratio tests and the Fisher distance.
We anticipate that the confidence intervals in $\boldg$ will continue 
to be based on likelihood ratio tests.
While both are invariant to reparametrization of $\boldg$, 
 non-linear terms in $\boldg$ that lead to curvature in the information
geometry can break the one-to-one relationship between the 
expected value of the log-likelihood ratio and the Fisher information distance.

In Fig.~\ref{fig:wbf_tautau_llr} we compare the two tools. As an
example, we study WBF Higgs production in the $\tau \tau$ mode and
sample parameter points $\boldg$ in the $\ope{W}$-$\ope{WW}$
plane. For each of these points we calculate the local and global
distance from the SM defined by the Fisher information, as well as the
expected log-likelihood ratio 
\begin{align}
  q(\boldg_b , \boldg_a )
  \equiv -2 \, E \left[
  \log \frac { f(\mathbf{x}|\boldg_b) }   { f(\mathbf{x} |\boldg_a) }
  \middle | \boldg_a \right] \,.
\end{align}
For small significance deviations, the local and especially the global distances are 
almost exactly equal to the expected likelihood ratio, 
with differences only becoming visible around the $3 \sigma$ level. 
This demonstrates that different
statistical tools probe the same physics and can be chosen based on
convenience.

\begin{figure}
  \includegraphics[height=0.45 \textwidth]{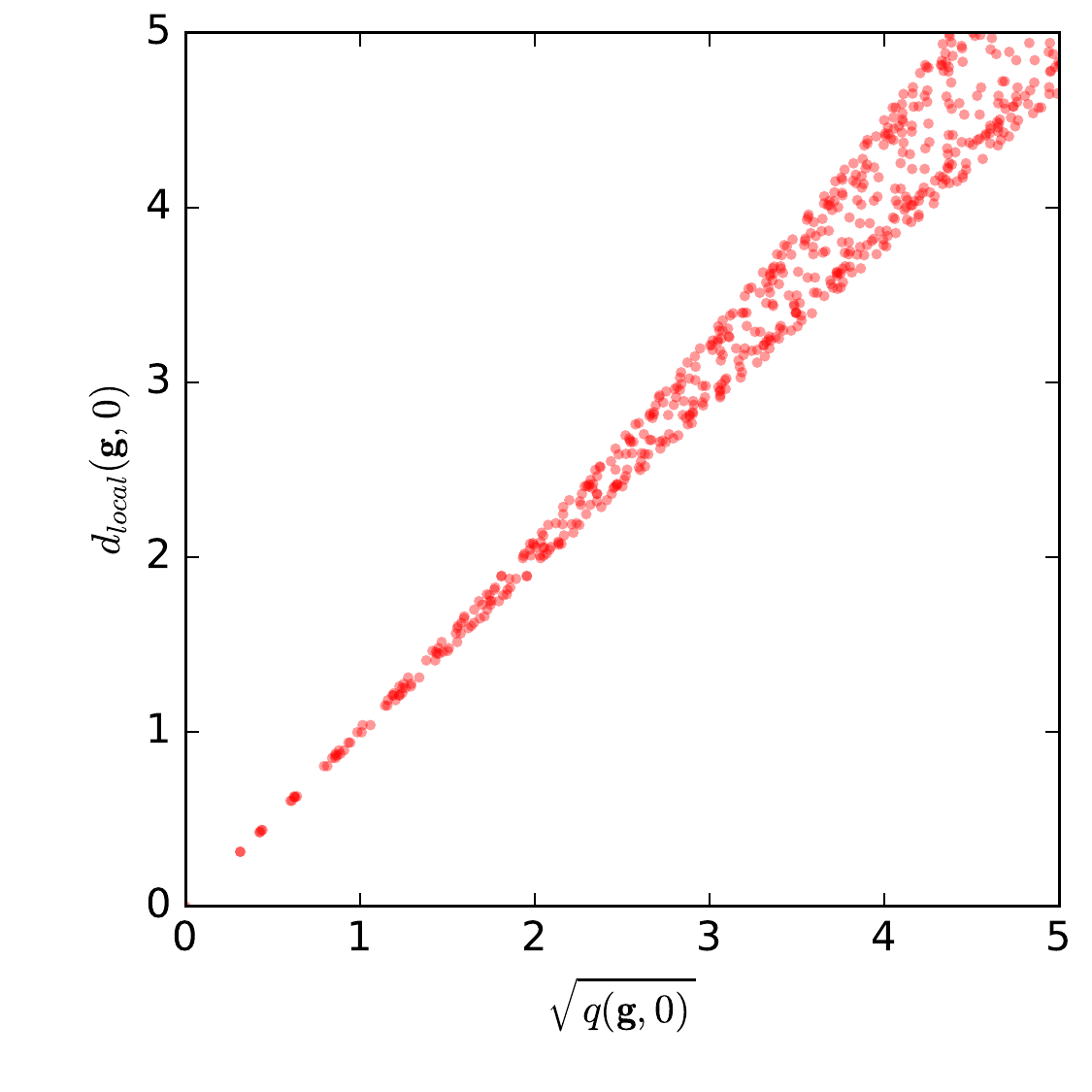}
  \hspace*{0.05\textwidth}
  \includegraphics[height=0.45 \textwidth]{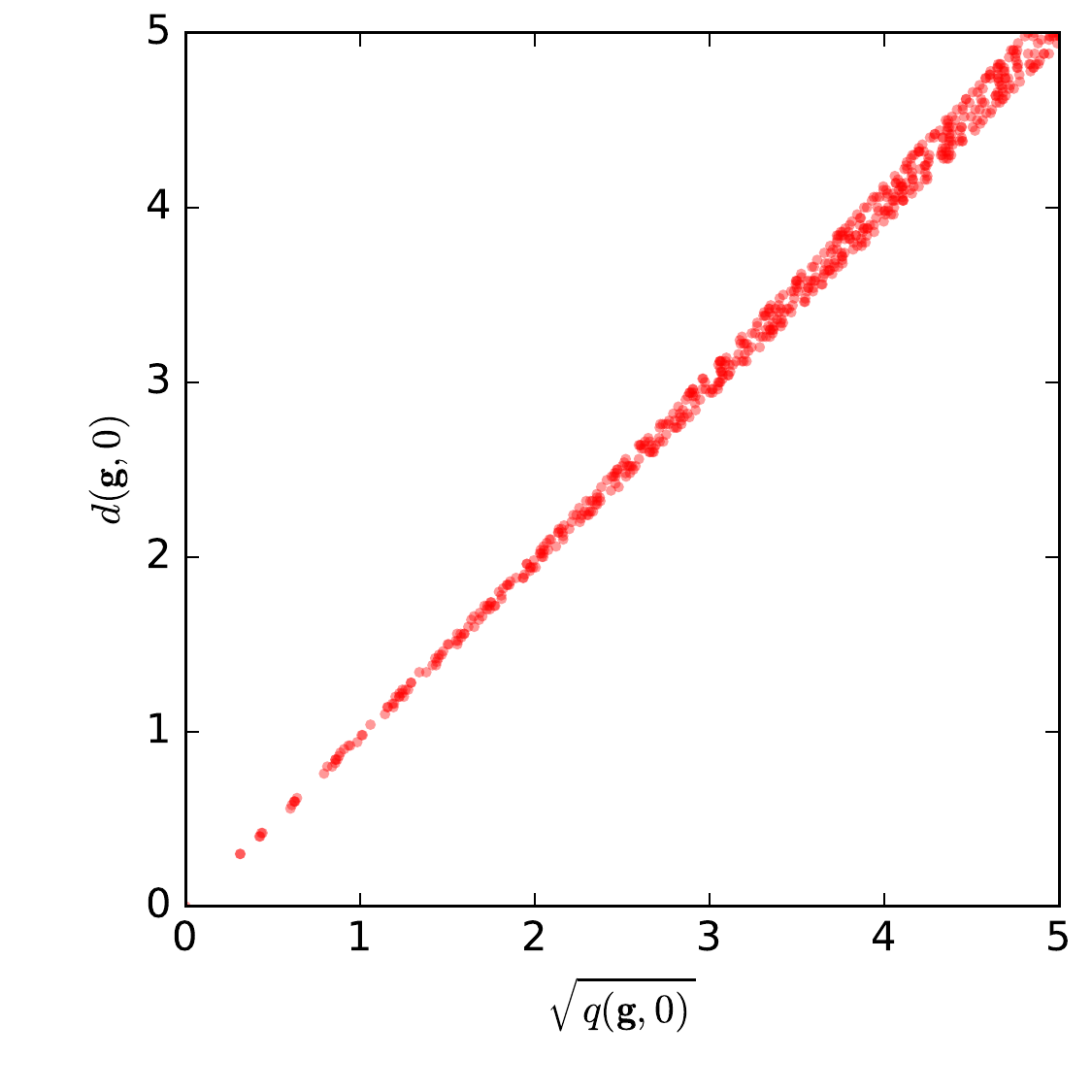}
  \caption{Comparison of the local (left) and global (right) distances
    defined by the Fisher information with the expected local
    likelihood ratio. We use WBF Higgs production in the $\tau \tau$
    mode and sample parameter points in the $\ope{W}$-$\ope{WW}$
    plane.}
  \label{fig:wbf_tautau_llr}
\end{figure}


\end{document}